# A Deep Learning Earth System Model Simulation of Indian Monsoon Intraseasonal and Interannual Variability

Bijit Kumar Banerjee[1,2], Devabrat Sharma[3,4], R. I. Sujith[3,4], Chandrashekar Lakshminarayanan[5], Manikandan Narayanan[5], Subodh K. Saha[6], Anurag Dipankar[7], Utpal Sarma[1,2], B. N. Goswami[1]

[1]ST Radar Centre, Gauhati University, India

[2]Department of Instrumentation and USIC

[3]Department of Aerospace Engineering, Indian Institute of Technology Madras, Chennai-600036, India

[4]Centre of Excellence for Studying Critical Transitions in Complex Systems, Indian Institute of Technology Madras, Chennai-600036, India.

[5]Department of Computer Science and Engineering, Indian Institute of Technology Madras, Chennai, India

[6]Indian Institute of Tropical Meteorology, Pune, India

[7]Institute for Atmospheric and Climate Science, ETH Zürich

**Abstract**

With the data-driven artificial intelligence / machine learning (AI/ML) models having demonstrated their ability to extend the prediction horizon of large-scale weather at a fraction of computational cost of numerical weather prediction models, a pertinent question is, could these models do the same for sub-seasonal to seasonal (S2S) prediction? A key challenge in developing a S2S prediction system is the requirement for a coupled ocean-atmosphere Earth system emulator that can stably simulate the observed intraseasonal and interannual variability with fidelity. In the rapidly evolving field of AI/ML weather models, such a deep learning 3D ocean-atmosphere coupled model has become available, called SamudrACE. With our interest in developing an AI/ML S2S model for Indian monsoon, here we examine the extent to which SamudrACE faithfully simulates Indian monsoon intraseasonal and interannual variability. Compared to observation, we found biases in SamudrACE's simulation of monsoon intraseasonal and interannual variability. Our systematic documentation and analyses of these biases provide a useful benchmark for improving not only SamudrACE but also coupled emulators in general and could fast track the development of a deep learning 3D global S2S prediction system.

## 1. Introduction

The monsoon intraseasonal variability is a key building block of the Indian monsoon system (Goswami, 2012; Goswami et al., 2006) through clustering the synoptic low-pressure systems (LPS) at one end and contributing to the interannual variability of the seasonal mean rainfall (Goswami & Mohan, 2001; S. K. Saha et al., 2019) at the other end of the spectrum. Improvements of prediction skill of weather and monsoon intraseasonal oscillations (MISO) (Goswami & Mohan, 2001) in the country has resulted in considerable benefit to the user community (farmers and fishermen) (Venkatesan et al, 2020). Prediction of both the MISO and the seasonal mean Indian summer monsoon rainfall (ISMR) require a coupled Ocean-Atmosphere model (Wang et al., 2005) and with progress in coupled modelling and data assimilation we can now predict the active and break spells of MISO about two weeks in advance (Sahai et al., 2013) while the limit on potential predictability may be about four weeks (Goswami et al., 2003; Lee & Wang, 2016; Waliser et al., 2003). Thus, there is considerable scope of improving the skill of MISO forecasts. However, the skill of forecasting MISO or Boreal Summer Intraseasonal Oscillations (BSISO) by the sub-seasonal to seasonal forecast (S2S) systems, has stagnated during the past decade (Fang et al., 2019). Similarly, while the potential skill of JJAS mean ISMR for 1-month lead forecasts may be 0.82 (M. Saha et al., 2016), the skill of most seasonal forecast systems is close to 0.6 (Rao et al., 2019). Further, a recent study (Sharma et al., 2022) has demonstrated that ISMR has high predictability even at 18-month lead while the skill of multi-model ensemble predictions by dynamical models decreases rapidly with lead time making forecasts useless after three months (Sharma et al., 2025). Thus, both the predictions of MISO and that of the seasonal mean ISMR by Atmosphere Ocean Coupled- General Circulation Models (AOGCMs) have reached a crossroad.

Like the weather, the limit on deterministic predictability of sub-seasonal oscillations or seasonal climate in dynamical prediction framework are also limited by growth of small errors in respective time scales. Since the first successful weather forecasts in 1950 (Charney et al., 1950; Charney & Phillips, 1953), weather forecasts have come a long way through a quiet revolution (Bauer et al., 2015) in improving the resolution and representation of physical processes in models and by reducing initial error with improved observations and data assimilation. However, weather predictions also faced a crossroad in recent decades where further improvement of skill of forecasts has become challenging and expensive. Therefore, the arrival of a new revolution in weather forecasting using artificial intelligence and machine learning (AI/ML) to make three-dimensional weather forecasts may be considered the beginning of the Second Revolution in Weather Forecasting (Bi et al., 2023; Kochkov et al., 2024; Lam et al., 2023; Pathak et al., 2022; Price & Rind, 1992). These different types of three-dimensional neural network models trained on high resolution (25 km x 25 km) hourly reanalysis (ERA5 (Hersbach et al., 2020)) have demonstrated that physically constrained data driven models can make weather forecasts for most large-scale variables as good as or better than the best numerical weather prediction (NWP) model at a small fraction of computation cost in an expanded prediction horizon. However, ERA5 hourly precipitation analysis has some deficiency compared to the observed hourly precipitation making it not suitable for training for hourly precipitation. As a result, initial AI/ML weather prediction models could not forecast precipitation even on a large scale. Precipitation has small-scale high frequency variability and requires sufficiently long records (like that of ERA5) of consistent circulation and precipitation from either reanalysis with a non-hydrostatic forecast model or from long simulations from a non-hydrostatic high-resolution weather/climate model (e.g. ICON) (Pothapakula et al., 2026; Prein et al., 2026). As precipitation is a positive only variable, without stringent physical constraints in training for precipitation, forecast model may produce unphysical negative precipitation. Rapid advances are being made in this direction, an example of which may be found in the latest version of European Centre for Medium-range Weather Forecasts (ECMWF) operational AI/ML model (AIFS) (Moldovan et al., 2025).

While rapid advances in AI/ML models for weather forecasting have been taking place during the past three years, development of an ocean-atmosphere coupled with AI/ML Earth system model (ESM) for S2S simulation (and prediction) has been lacking for several reasons. Firstly, training data requirement for S2S models is at least an order of magnitude larger than that for weather prediction and simulation and such observational (or reanalysis) data is scarce. Secondly, the coupling between the ocean and atmosphere in the AI/ML models should be able to simulate physically meaningful processes (e.g. Kelvin and Rossby waves in the Atmosphere and the Ocean) that are responsible for evolution of the climate system. It is notable that one such model has just been developed, named SamudrACE (Duncan et al., 2026). SamudrACE is an AI emulator of the Geophysical Fluid Dynamics Laboratory (GFDL) CM4 (Dunne et al., 2020) coupled global climate model (GCM). Different components of the coupled emulator are developed separately and trained on 150 years of pre-industrial output. The coupled emulator is computationally efficient and can produce 1500 years of simulations in one day on a single NVIDIA H100 GPU (see Data and Methods for details). (Duncan et al., 2026) presents the first successful and stable AI Earth System Model (ESM) that simulates the observed mean global climate and its short-term climate variability within reasonable bounds. We consider this to be a milestone in data driven climate emulators and provide an opportunity to develop a physically meaningful data driven S2S prediction system. We are particularly interested in developing a S2S system for the Indian monsoon intraseasonal and interannual variability. For this purpose, it is imperative to assess the ability of the AI model in simulating the observed MISO and interannual variability of Indian summer monsoon rainfall (ISMR) well. Towards that end, we estimate the biases in simulating the observed MISO as well as in simulating the interannual variability of ISMR. Unfortunately, daily data from the original simulation was not available. Therefore, we ran the SamudrACE at Gauhati University, India for 300 years and saved the daily fields. Here, we report preliminary results of analysis of monsoon intraseasonal, and interannual variability simulated by the AI coupled model and try to assess the strengths and weaknesses of the model. We believe that this is a prerequisite and useful step in building a cost effective S2S prediction system for Indian monsoon enhancing both the skill and prediction horizon.

## 2. Data and Methods:

### 2.1 SamudrACE:

To study the intraseasonal and interannual variability of the Indian summer monsoon, we use the AI-based coupled Earth system model SamudrACE (Duncan et al., 2026). It is a data-driven emulator of a fully coupled atmosphere–ocean model trained on long simulations from the Geophysical Fluid Dynamics Laboratory Climate Model version 4 (GFDL-CM4) (Dunne et al., 2020). Instead of explicitly solving physical equations, the model learns the evolution of the system using neural networks. This allows much faster simulations while still maintaining realistic large-scale behavior.

The model has two components: an atmospheric model (ACE2) (Watt-Meyer et al., 2023) and an ocean model (Samudra) (Dheeshjith et al., 2025). These are trained separately and then coupled through exchange of surface fluxes such as heat, momentum, and freshwater. The coupling is physically constrained, where the atmosphere forces the ocean through surface fluxes, while the ocean feeds back through sea surface temperature and sea ice.

The atmospheric and oceanic components are implemented as three-dimensional deep neural networks that emulate the evolution of the corresponding state variables. Together, the coupled system contains approximately 600 million trainable parameters. Both components are first pretrained independently using output from CM4 and are then jointly fine-tuned in coupled mode to ensure stable interaction between the atmosphere and ocean. This modular architecture closely follows the structure of conventional coupled climate models while retaining the computational efficiency of machine learning-based emulators.

A key aspect of the model is the difference in time stepping between the two components. The atmosphere evolves at a 6-hour time step, while the ocean is updated every 5 days. During each 5-day period, the atmosphere is integrated forward in 6-hour steps, and the resulting surface fluxes are averaged over this period. These 5-day mean fluxes are then used to force the ocean model, which advances one step to update the ocean state. The updated sea surface temperature is then passed back to the atmosphere, and the coupling cycle continues. This setup reflects the faster evolution of the atmosphere compared to the ocean and is important for maintaining stable coupled variability.

The model runs at about 100 km resolution and simulates key variables such as winds, temperature, salinity, and surface fluxes. The atmosphere captures fast processes like convection and circulation, while the ocean represents slower changes in heat content and stratification. Together, they produce variability across timescales relevant to the monsoon.

An important advantage of SamudrACE is its computational efficiency, which allows very long climate simulations to be performed at a small fraction of the computational cost of traditional coupled climate models. In this study, we conducted three independent 300-year simulations, each initialized from a different synthetic year of the NOAA Geophysical Fluid Dynamics Laboratory CM4 preindustrial simulation used to train SamudrACE, together with the corresponding model state and forcing fields. These synthetic years are indices within the continuous CM4 simulation and do not correspond to actual calendar years. For example, "year 311" refers to the 311th year of the CM4 preindustrial control simulation. Starting from the CM4 state and forcing associated with this synthetic year, SamudrACE was integrated forward for an additional 300 years. Most of the analyses presented in this study are based on this simulation because daily output was archived throughout the integration, enabling a detailed examination of both intraseasonal variabilities, including active and break cycles, and interannual variability of the Indian summer monsoon.

**2.2 Data:**

For verification of the simulated monsoon variability, we use multiple observational and reanalysis datasets. Intraseasonal variability of rainfall, including equatorial wave characteristics, is evaluated using the Climate Prediction Center Morphing Technique data set (CMORPH) (Pingping Xie et al., 2018; Xie et al., 2017), which provides high-resolution satellite-based daily precipitation estimates suitable for capturing wave-scale variability. To ensure consistency with the resolution of SamudrACE and CM4, the high resolution 8 km CMORPH data area averaged to make ~100km resolution. For interannual variability of Indian summer monsoon rainfall, we use the IITM Parthasarathy RR-65 data set (Parthasarathy et al., 1994), a long-term record (1871-2016) of rainfall over Indian landmass. To examine sea surface temperature variability and its relationship with monsoon and ENSO, monthly SST from the COBE SST dataset (Schneider et al., 2013) is used, along with daily SST from ERA5 (Hersbach et al., 2020) for higher temporal resolution analysis. In addition, 10 m zonal wind (u10) from ERA5 is used to examine low-level wind variability and ocean–atmosphere interactions.

**2.3 Methods**

Intraseasonal variability was analyzed using diagnostics of equatorial wave characteristics and monsoon propagation. Convectively coupled equatorial waves were identified using Wheeler–Kiladis space–time spectral analysis (Hendon & Wheeler, 2008; Wheeler & Kiladis, 1999) of daily precipitation over the tropical belt (±15°). A sliding window of approximately 96 days was applied to compute frequency–zonal wave number spectra and to isolate dominant wave modes relative to theoretical dispersion relationship (Hayashi, 1982; Matsuno, 1966). The northward propagation of intraseasonal variability over the Indian monsoon region was further examined using frequency–meridional wavenumber spectra computed over 60°–110°E and 20°S–40°N. Rainfall anomalies were detrended, weighted by latitude, and decomposed using two-dimensional Fourier transforms, with statistical significance assessed against a red-noise background (Hendon & Wheeler, 2008).

The dominant monsoon intraseasonal oscillation (MISO) was extracted using extended empirical orthogonal function (EEOF) analysis applied to daily rainfall anomalies over the core monsoon domain (60.5°–95.5°E, 12.5°S–30.5°N). Temporal embedding of approximately 15 days was used to capture propagating structures. The leading two principal components were normalized to define the MISO index and its amplitude (Suhas et al., 2013). Their characteristics were further examined using power spectra, lag–lead correlations, and phase composite analysis based on eight standard MISO phases to describe the spatio-temporal evolution of the oscillation.

Interannual variability of monsoon rainfall was characterized using EOF analysis of JJAS-mean rainfall anomalies, with latitude-based area weighting applied to account for spherical geometry. The leading modes were used to diagnose dominant spatial patterns. Sea surface temperature variability was analyzed using the Niño-3.4 index, defined as the area-weighted SST anomaly over 5°S–5°N and 190°–240°E after removal of the seasonal cycle, along with EOF analysis of tropical SST anomalies (±30°) to identify ENSO-related modes. The coupling between ENSO and the Indian monsoon was quantified using simultaneous, running, and lag–lead correlations between SST indices and JJAS rainfall anomalies to assess the strength and phase relationship of the teleconnection.

## 3. Results

### 3.1 Atmospheric Equatorial wave

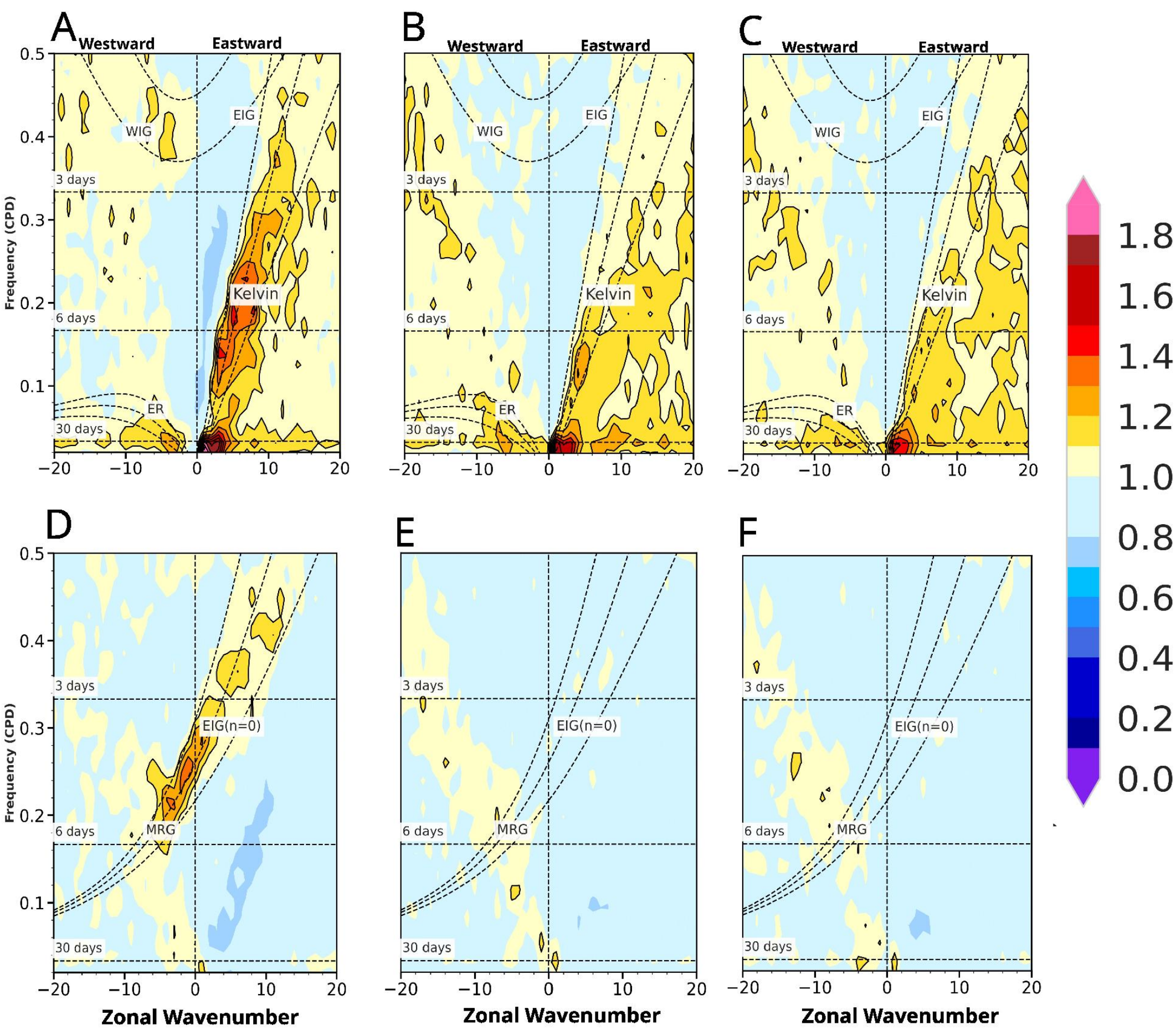

*Figure1 Wheeler–Kiladis space–time spectra of daily accumulated precipitation computed over a 25-year period for (A) CMORPH, (B) CM4, and (C) SamudrACE. Panels (A–C) show the symmetric component, while panels (D–F) display the antisymmetric component. The spectra are overlaid with Matsuno dispersion curves representing equatorial wave modes.*

As the convectively coupled equatorial waves (CCEW) and the seasonal mean climate form an interacting system in quasi equilibrium (Gill, 1980), biases in simulating the mean may be related to biases in simulating the CCEW. Since the intraseasonal oscillations and the seasonal mean are interdependent, biases in the mean would introduce biases in simulations of intra-seasonal variability. Therefore, we examine the simulation of the CCEW by SamudrACE by calculating the space time spectra following Wheeler and Kiladis (1999) . For this purpose, we use daily precipitation data for the full year from 30 years of SamudrACE and CM4 simulations, and 25 years of satellite precipitation derived from CMORPH (Xie et al., 2017). It is notable that the emulator simulates most of the symmetric waves (MJO and westward propagating long Rossby waves, westward propagating quasi biweekly mode, and the eastward propagating Kelvin waves) reasonably well (Figure 1A, B, C). However, CM4 underestimates eastward propagating atmospheric Kelvin waves with periods between 3 and 10 days (Figure 1B), which SamudrACE further weakens (Figure 1C), and remain severely underestimated. A significant bias is also noted in simulating the antisymmetric waves (Figure 1D, E, F). The model underestimates long mixed Rossby gravity (MRG) waves and also underestimates the shorter westward propagating MRG waves (between global wave numbers -2 and -5). This bias is also present in CM4 and seems to arise from deficiencies in the training data. Weak atmospheric equatorial waves may lead to some weak oceanic equatorial waves and in turn may lead to weak air-sea interactions over the tropics affecting the simulated intraseasonal and interannual variability.

The northward propagating monsoon intraseasonal oscillation (MISO) is a regional phenomenon during boreal summer. To examine the simulation of the MISO, we construct regional space time spectra for north south propagating waves. We average daily precipitation between 70°E and 100°E and compute space time spectra, where space is represented by latitudes between 20°S and 40°N, using the same years as in Figure 1. It is notable that the observations (Figure 2A) are dominated by a northward propagating wave with a period of about 40 days and meridional wave numbers 1 to 2 (between 20°S and 40°N). However, CM4 simulates a dominant northward propagating mode with a 60-day period and wave number 2, with little power at the 40-day period (Figure 2B). SamudrACE carries forward this bias (Figure 2C) while introducing a stronger stationary component with a period of 60 days.

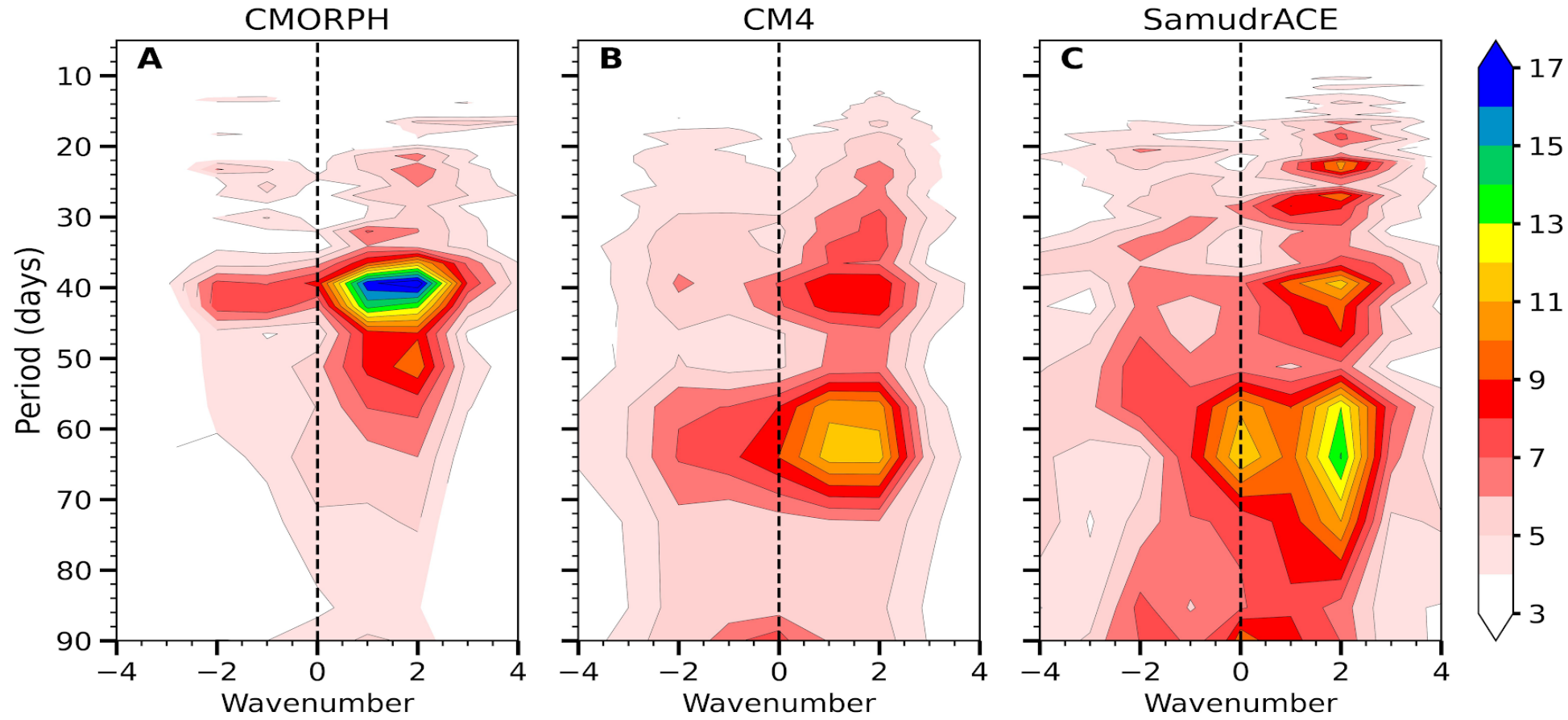

*Figure2 Meridional wavenumber–frequency spectra of JJAS precipitation over $20^0$ S–30°N, 65.5–100.5°E for CMORPH, CM4, and SamudrACE, showing longitude-averaged and latitude-weighted variability. Shading indicates red-noise–normalized power with only 90% significant signals retained, highlighting dominant intraseasonal variability in the 2–90-day band.*

### 3.2 Monsoon Intraseasonal Oscillations (MISO)

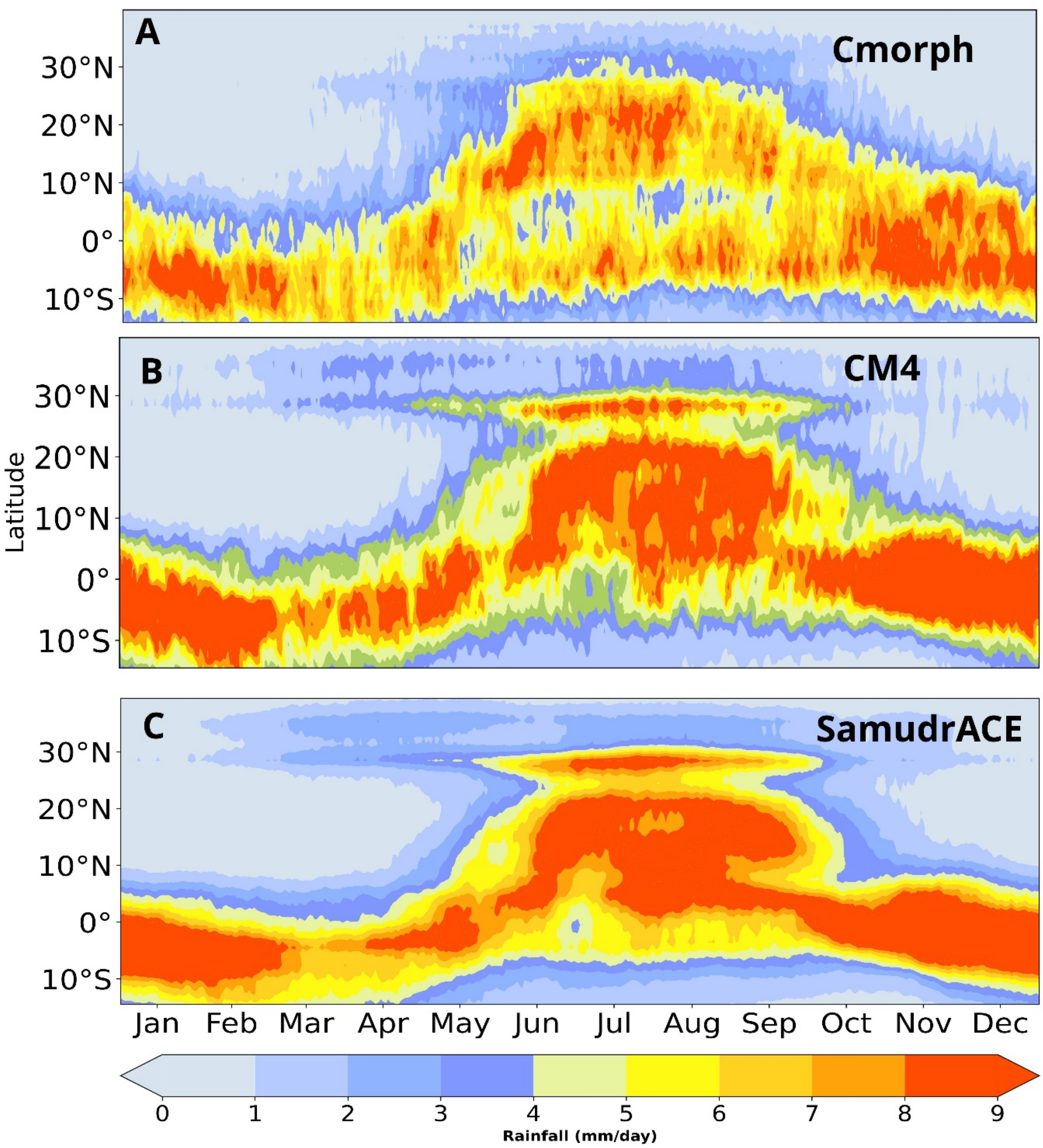

*Figure3. Annual cycle of the northward migration of the rain band. (A) Climatological daily rainfall averaged over 70°E–90°E as a function of latitude from observations (CMORPH). (B)Same as (A), but from the 300-year CM4 simulation. (C) Same as (A) but simulated by SamudrACE.*

The MISO is a manifestation of fluctuations of the intertropical convergence zone (ITCZ) (Goswami, 2012; Goswami & Shukla, 1984) on intraseasonal time scales, and any bias in simulating the ITCZ over the Indian monsoon region by SamudrACE would be reflected in its simulation of the MISO. Therefore, we examine the simulated climatological annual cycle of the rain band (ITCZ) by SamudrACE (Figure 3C) and compare it with observations (Figure 3A). In contrast to most CMIP6 models, in which the northward migration of the rain band is limited to between 13° and 15°N (Choudhury et al., 2022), SamudrACE simulates the northward migration of the rain band up to approximately 22°N, consistent with observations (Figure 3A). The main bias with potential implications for the MISO is that the secondary oceanic rain band south of the continent is simulated by SamudrACE north of the equator (approximately 5°N), whereas it is located south of the equator (approximately 5°S) in observations. A comparison with CM4 (Figure 3B) shows a similar bias, with the oceanic ITCZ positioned at or north of the equator. Therefore, SamudrACE inherits this bias from its training

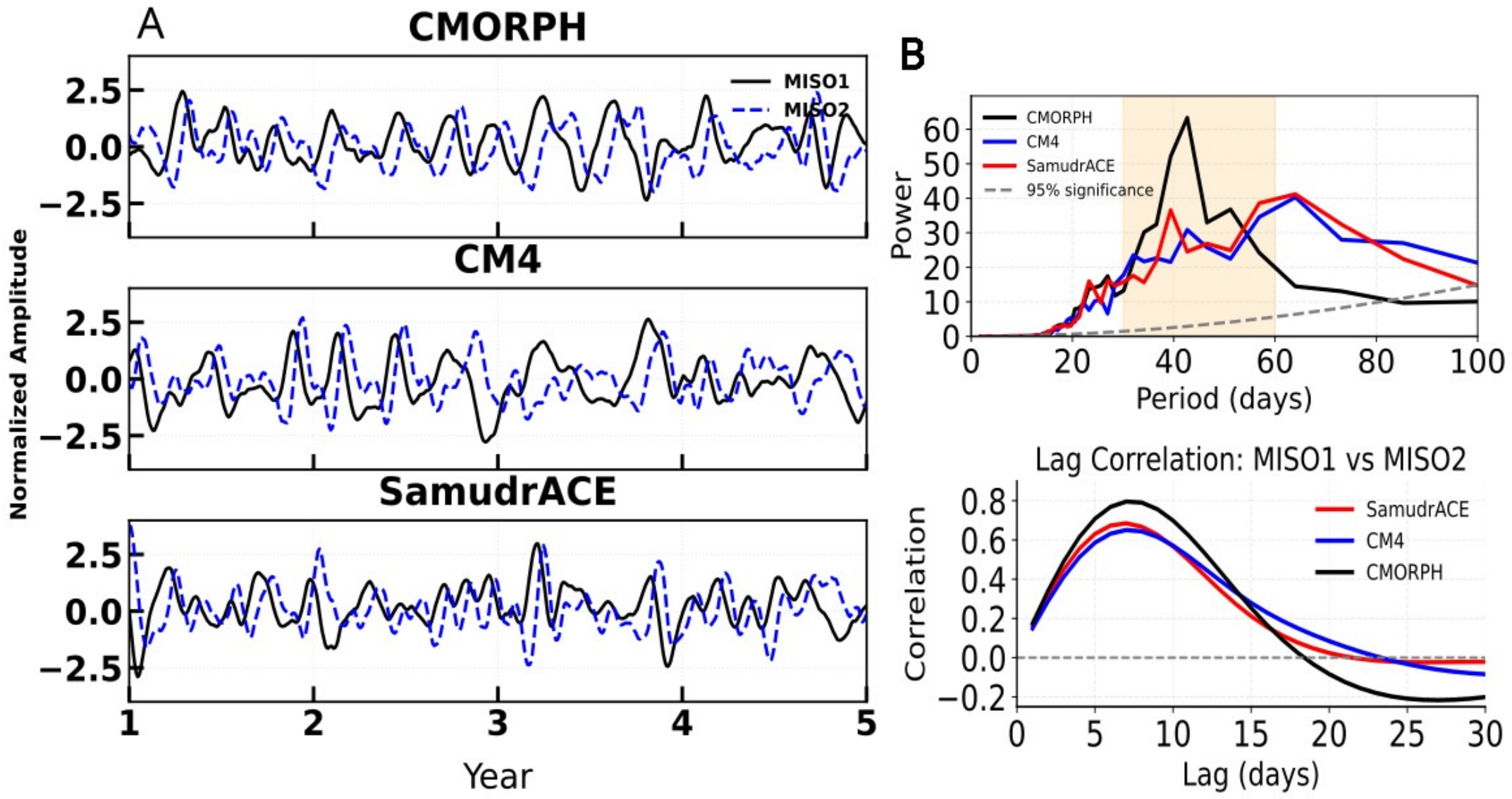

*Figure4 (A) The first and second principal components (MISO1 and MISO2) of the extended EOF analysis derived from the datasets normalized by their own standard deviation. (B ) Power spectrum of MISO1 from SamudrACE, CM4 and observations. (C) The lag correlation between MISO1 and MISO2 up to 30 days.*

data.

To compare the simulation of the dominant mode of intraseasonal variability during Boreal summer, we use a real time monitoring index of the MISO (Suhas et al., 2013), represented by the two leading principal components (MISO1 and MISO2) from an extended EOF analysis of rainfall averaged over 60.5°E to 95.5°E over the Indian monsoon region, for SamudrACE and CM4, and compare these with observations (Figure 4A). The lead lag correlation between MISO1 and MISO2 in observations indicates a northward propagation of the rain band from south of the equator to the sub-Himalayan region of northern India in about 35 days. Both CM4

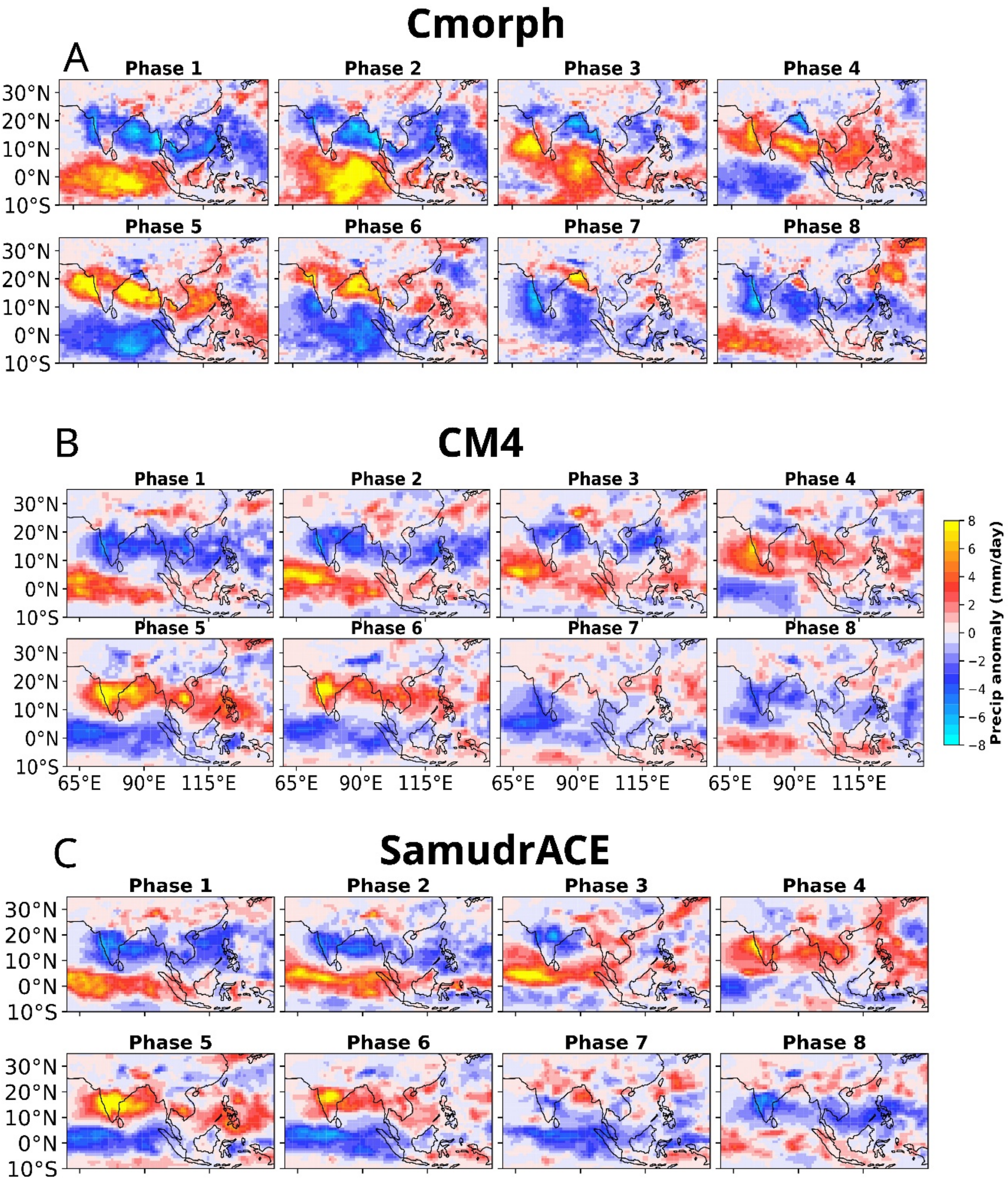

*Figure5 Spatial distribution of composite rainfall anomalies corresponding to the eight phases of the MISO. (A) Observations from CMORPH, (B) CM4 simulations, and (C) SamudrACE simulations. Following the evolution of the MISO each year (as in Figure. 4), each day is assigned to one of the eight phases, and composite daily rainfall anomalies are computed using all days within each phase.*

simulate the propagation speed reasonably well, but the amplitude of the rain band is weaker than observed. Consistent with Figure 2, the power spectrum of MISO1 from SamudrACE, CM4, and observations (Figure 4B) shows a clear bias. Observations indicate maximum power at a period of 42 days, whereas both CM4 and SamudrACE simulate maximum power at a period of 60 days. The power at 42 days is about 50% of that observed, while the power at 60 days is three to four times larger than observed. This is a major bias and may limit the model's usefulness as an S2S prediction system. However, this bias could potentially be reduced through retraining with output from a higher resolution Earth system model with improved MISO representation.

Following Suhas et al. (2013) the evolution of the MISO in the two-dimensional phase space defined by MISO1 and MISO2 is illustrated for two years from SamudrACE, CM4, and observations (Figure S2). Circular trajectories indicate the northward propagation of individual events. Large amplitude events are characterized by large radii, whereas smaller radii correspond to weaker events. Although the two simulated years are not directly comparable with the observational years, SamudrACE captures the event-to-event variability within a year reasonably well.

The composites (Figure 5) show the northward migration of the rain band as a function of phase in observations (CMORPH), SamudrACE, and CM4. The rain band is weaker in CM4 (Figure 5B) and in SamudrACE (Figure 5C) compared with observations (Figure 5A). In observations, as the rain band propagates northward, it becomes tilted, extending from southeast near the equator to northwest at higher latitudes. In contrast, both CM4 and SamudrACE maintain a largely zonal structure throughout the northward propagation. This is a significant bias and may further limit the suitability of SamudrACE as a candidate model for Indian monsoon S2S prediction.

**3.3 Indian monsoon Interannual variability**

The simulated climatological summer (JJAS) and winter (DJF) precipitation from observations (Figure S3A, B), SamudrACE (Figure S3E, F), and CM4 (Figure S3C, D) indicate that, while SamudrACE does not exhibit a double ITCZ bias over the equatorial Pacific as seen in many climate models, it overestimates the amplitude of the DJF ITCZ over the ocean. This overestimation of oceanic precipitation appears to originate from the parent Earth system model CM4 on which it is trained (Figure S3C). To highlight biases over the Indian monsoon region, the differences between JJAS and DJF climatology from SamudrACE, CM4, and observations are shown in Figure 6G, H and Figure 6I, J, respectively. A significant overestimation of the DJF ITCZ over the ocean is evident (Figure 6H, J) in both CM4 and SamudrACE. During JJAS, the artificial ITCZ between the equator and 10°N in SamudrACE (Figure 6G) leads to a tripole pattern in meridional direction of biases with overestimation between the equator and 15°N, underestimation between equator and 10°S and between 15°N and 25°N as noted in Figure S3G.

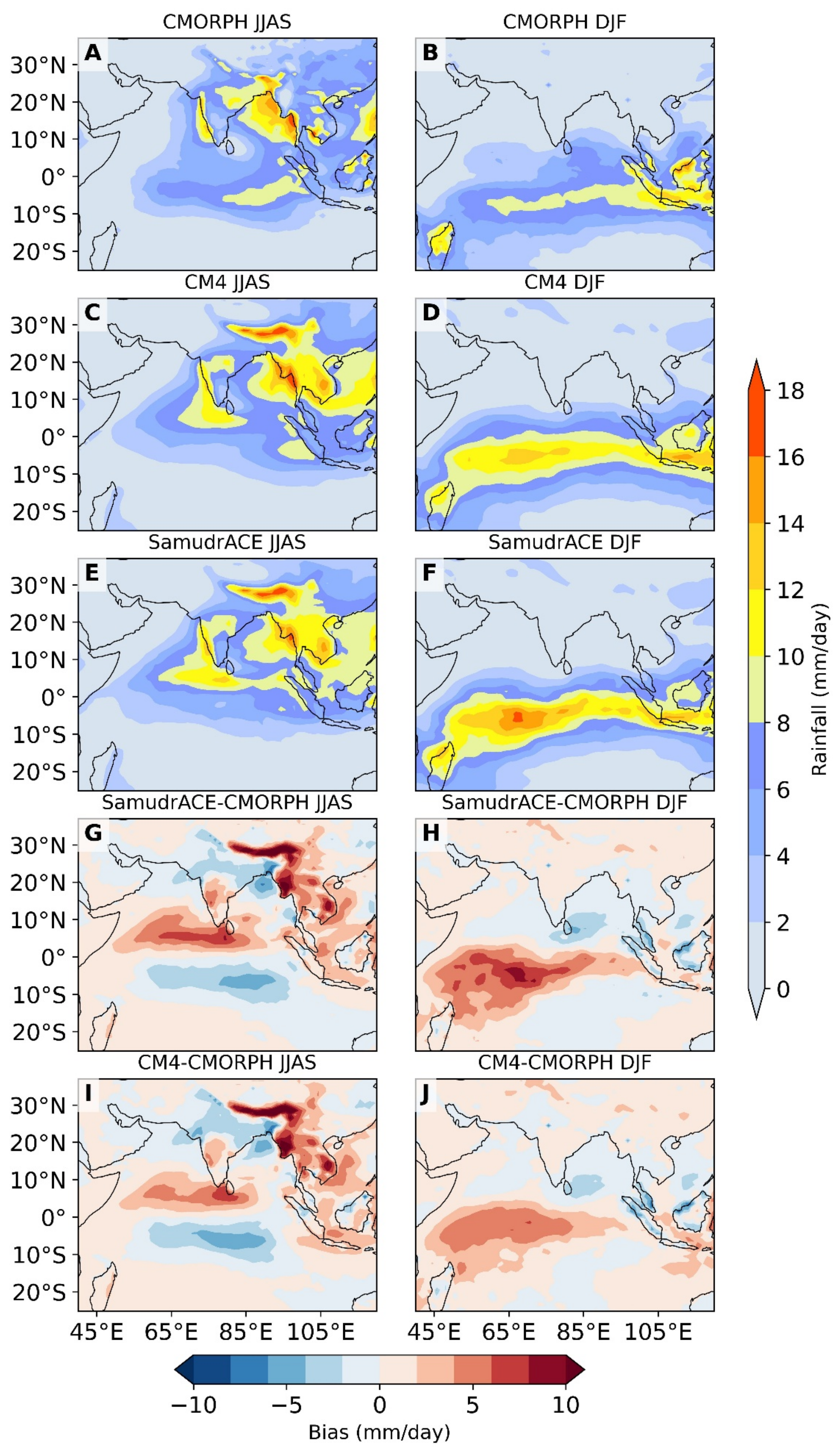

*Figure6: Climatological spatial structure of rainfall over the ISMR domain (5°–40°N, 60°–110°E) for the JJAS (left column) and DJF (right column) seasons. Panels (A, B) show CMORPH; (C, D) show observations from CM4 (100 km); and (E, F) show SamudrACE. Panels (G, H) present the difference between SamudrACE and CMORPH for JJAS and DJF, respectively, while panels (I, J) show the difference between CM4 and CMORPH for JJAS and DJF.*

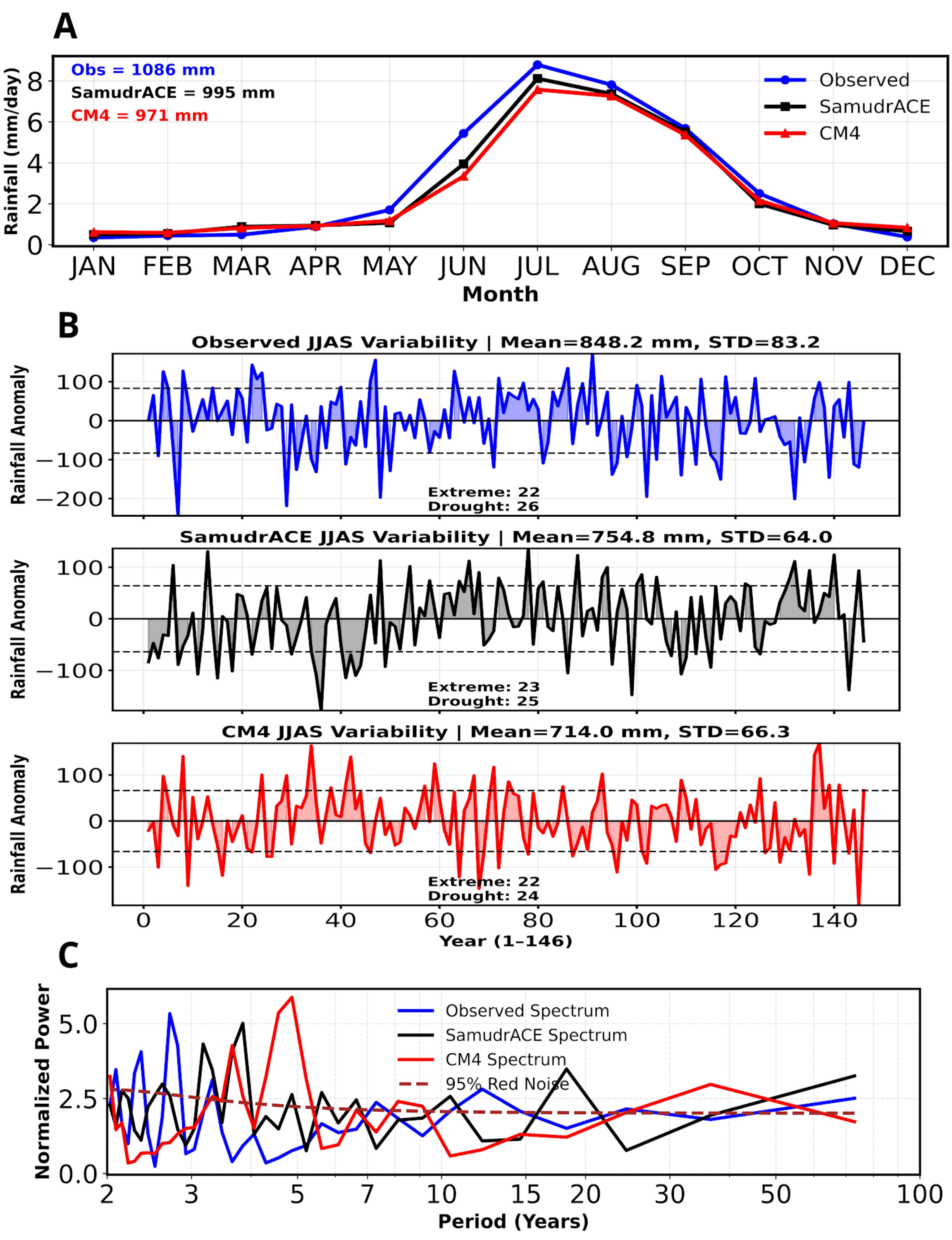

*Figure7. (A) Climatological annual cycle of rainfall averaged over the Indian landmass from IITM RR65 observations (blue), GFDL CM4 (red), and SamudrACE (black). (B) Time series of JJAS mean interannual rainfall anomalies over India from IITM RR65 observations (blue), CM4 (red), and SamudrACE (black). (C) Power spectra of JJAS interannual rainfall variability from IITM RR65 observations (blue), CM4 (red), and SamudrACE (black).*

compared with the observed 845 mm, corresponding to 89% of the observed long-term mean. The normalized time series of interannual variability of JJAS rainfall over India simulated by SamudrACE (Figure 7B) indicates that the amplitude of variability is substantially weaker than observed and is similar to that simulated by CM4. The standard deviation of interannual variations in JJAS mean rainfall is 64.0 mm in SamudrACE and 66.3 mm in CM4, compared with 83.2 mm in the observations (RR65). This similarity suggests that the weak interannual variability in SamudrACE is largely inherited from its training data. Considering JJAS rainfall anomalies greater than +1 standard deviation and less than −1 standard deviation as extreme wet and dry years, respectively, the observations contain 22 excess monsoon years and 24 deficient monsoon years, the SamudrACE also simulates a comparable number of extreme events over the analysis period. In the frequency domain, both observations and SamudrACE exhibit multidecadal variability with periods between 60 and 80 years. However, SamudrACE shows a pronounced decadal oscillation of nearly 20 years that is absent in the observations (Figure 7C). At shorter interannual timescales, the observations are dominated by a biennial component, whereas SamudrACE exhibits a dominant oscillation with a period of approximately four years.

### 3.3.1 Simulation of dominant mode of Interannual variability of JJAS rainfall over the region

Here, we examine how ERA5, GFDL-CM4, and SamudrACE represent the spatial patterns and amplitudes of the dominant modes of observed interannual variability in seasonal mean rainfall over India (Fig. 8). EOF1 patterns of JJAS rainfall over Indian land mass for SamudrACE (Fig. 8A), CM4 (Fig. 8B), and ERA5 (Fig. 8C) show strong agreement, indicating that both models realistically capture the leading mode of variability. The fraction of variance explained by EOF1 is also nearly identical across the three datasets, accounting for 23.6% in SamudrACE, 22.9% in CM4, and 23.5% in ERA5. This demonstrates that SamudrACE reproduces both the spatial structure and variance of the primary observed mode with fidelity comparable to that of its parent dynamical model, a notable positive for a seasonal prediction model.

For EOF2, SamudrACE (Fig. 8D) simulates the variance (13.5%) close to that by ERA5 (14.7%) closely following CM4 (13.8%). In contrast to a east-west dipole spatial pattern for the EOF1 (Fig.8A, B, C), the EOF2 is associated with a north-south tripole pattern SamudrACE (Fig.8D) inherits from CM4 (Fig.8E). Overall, SamudrACE simulates both the amplitude and spatial pattern of dominant interannual variability over the Indian continent well within tolerable uncertainty, again a positive for a seasonal prediction model for Indian monsoon. The principal component (PC) time series associated with EOF1 for SamudrACE, CM4, and ERA5 (Figs. 8G–I) exhibit very similar amplitudes and temporal variability, further confirming that SamudrACE reproduces the magnitude and evolution of the dominant interannual rainfall mode well.

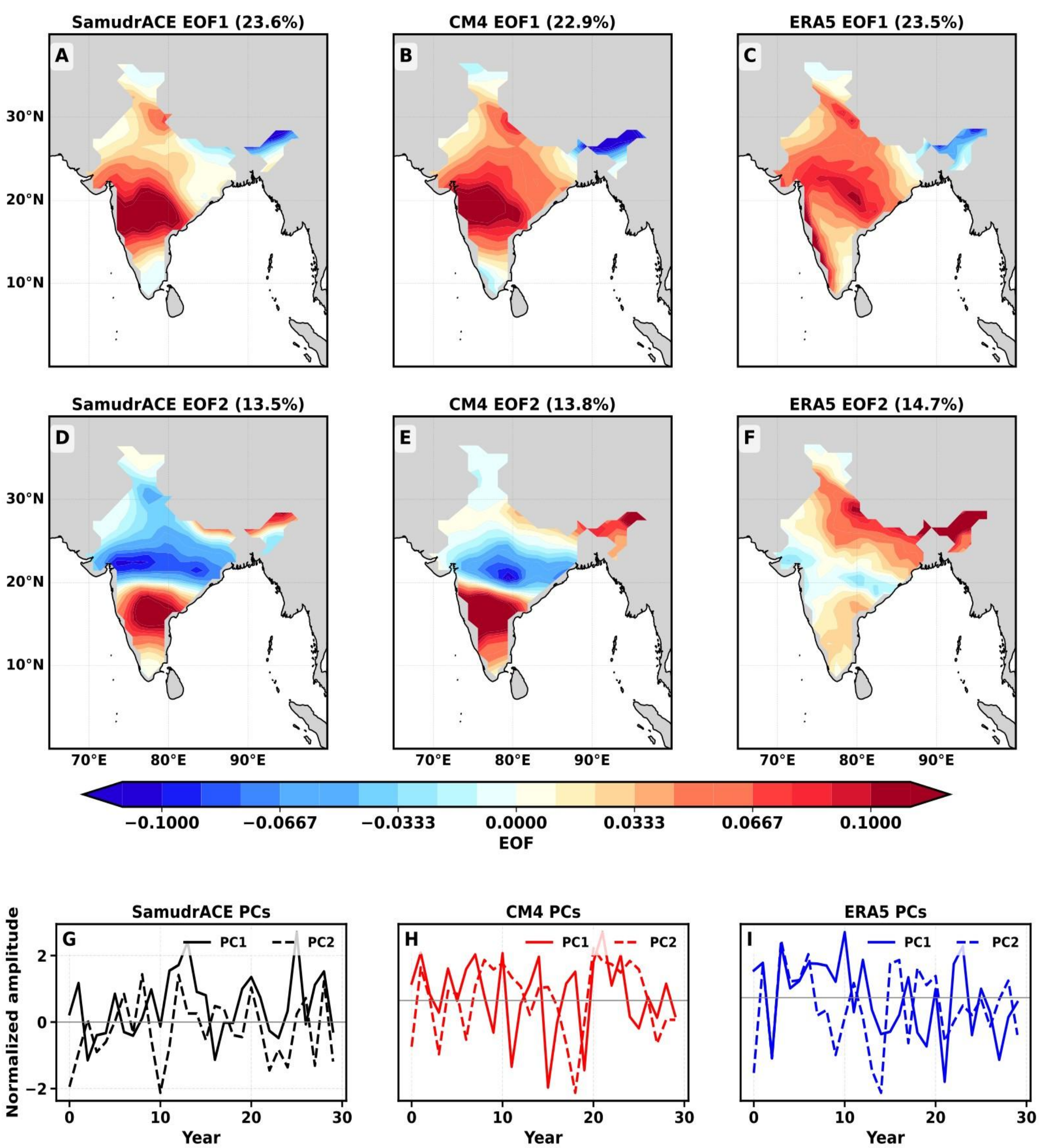

*Figure8. Leading empirical orthogonal function (EOF) modes of JJAS rainfall anomalies over the Indian monsoon region computed from 84 years of data. Panels (A) and (D) show EOF1 and EOF2 from SamudrACE, respectively. Panels (B) and (E) show the corresponding EOF1 and EOF2 from NOAA-GFDL CM4, and panels (C) and (F) show EOF1 and EOF2 derived from observations (ERA5) for 84 years. Panels (G), (H), and (I) present the standardized principal components (PC1 & PC2) time series associated with EOF1 and EOF2 for SamudrACE, CM4, and ERA5, respectively. EOF1 and EOF2 represent the dominant spatial patterns of interannual variability in JJAS rainfall.*

### 3.3.2 Simulation of ENSO-Monsoon Relationship

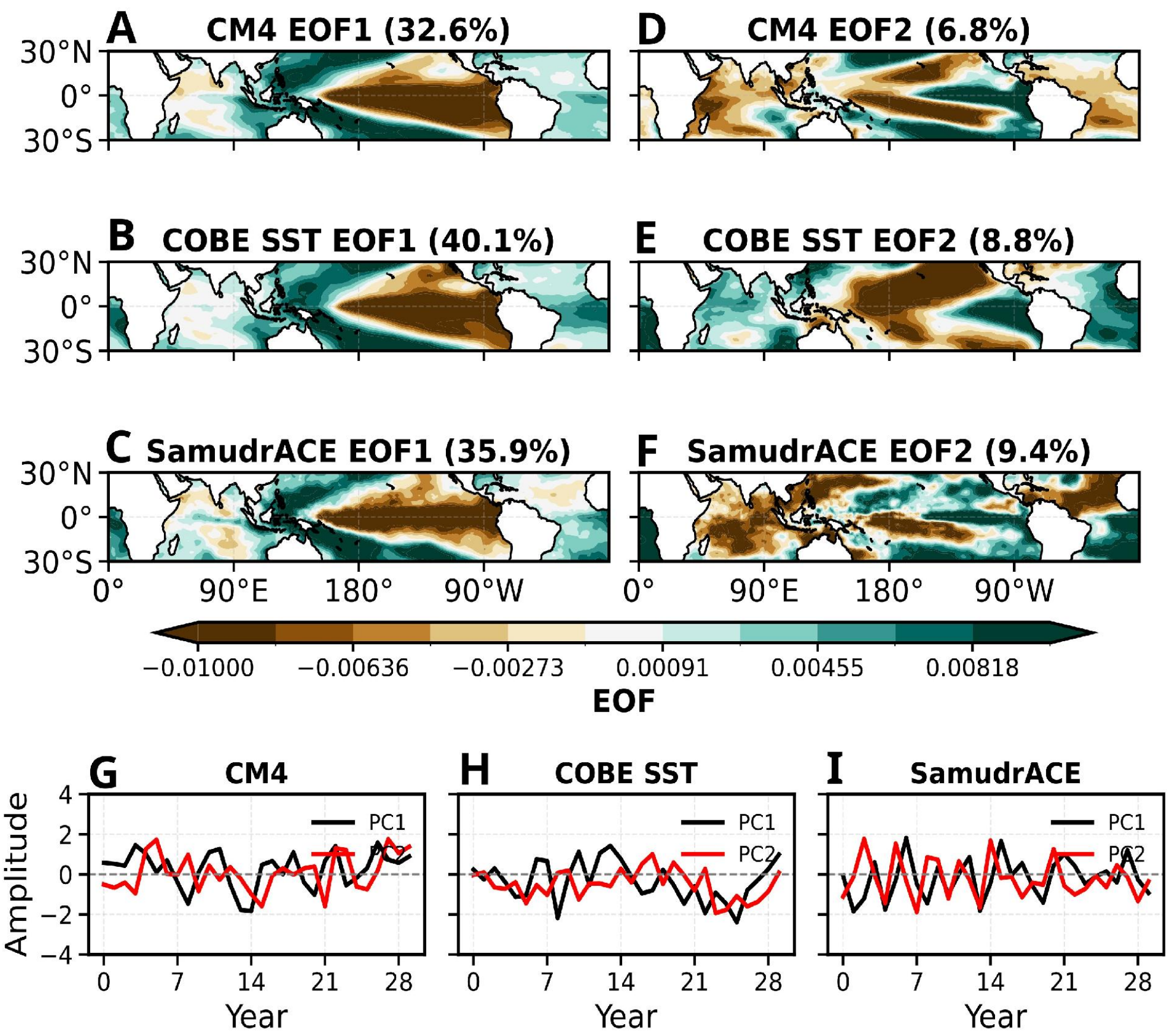

*Figure9. Leading empirical orthogonal function (EOF) modes of annual mean tropical (30°S–30°N) sea surface temperature (SST) anomalies after removal of the seasonal cycle and tropical mean, computed over 126 years. Panels (A) and (D) show EOF1 and EOF2 from NOAA GFDL CM4, respectively. Panels (B) and (E) show the corresponding EOF1 and EOF2 derived from observations using NOAA National Centers for Environmental Information COBE-SST, and panels (C) and (F) show EOF1 and EOF2 from SamudrACE. The percentage of variance explained by each mode is indicated in the panel titles. Panels (G), (H), and (I) present the standardized principal component (PC1 and PC2) time series corresponding to CM4, COBE-SST, and SamudrACE, respectively.*

Given that El Niño–Southern Oscillation (ENSO) is the dominant driver of ISMR variability, a realistic simulation of the observed ENSO–monsoon relationship by SamudrACE is a prerequisite for its use as a seasonal prediction model of ISMR. Therefore, we first examine the simulation of the dominant modes of sea surface temperature (SST) variability by SamudrACE and compare them with observations (Figure 9).

In terms of both the variance explained and the spatial pattern of the dominant mode (EOF1), which we term as the Global-ENSO mode (Fig.9B), SamudrACE performs well (Fig.9C). For EOF2, although the variance explained is comparable to observations (Fig.9E), the spatial pattern (Fig.9F) differs substantially from the observed structure but similar to EOF2 simulated

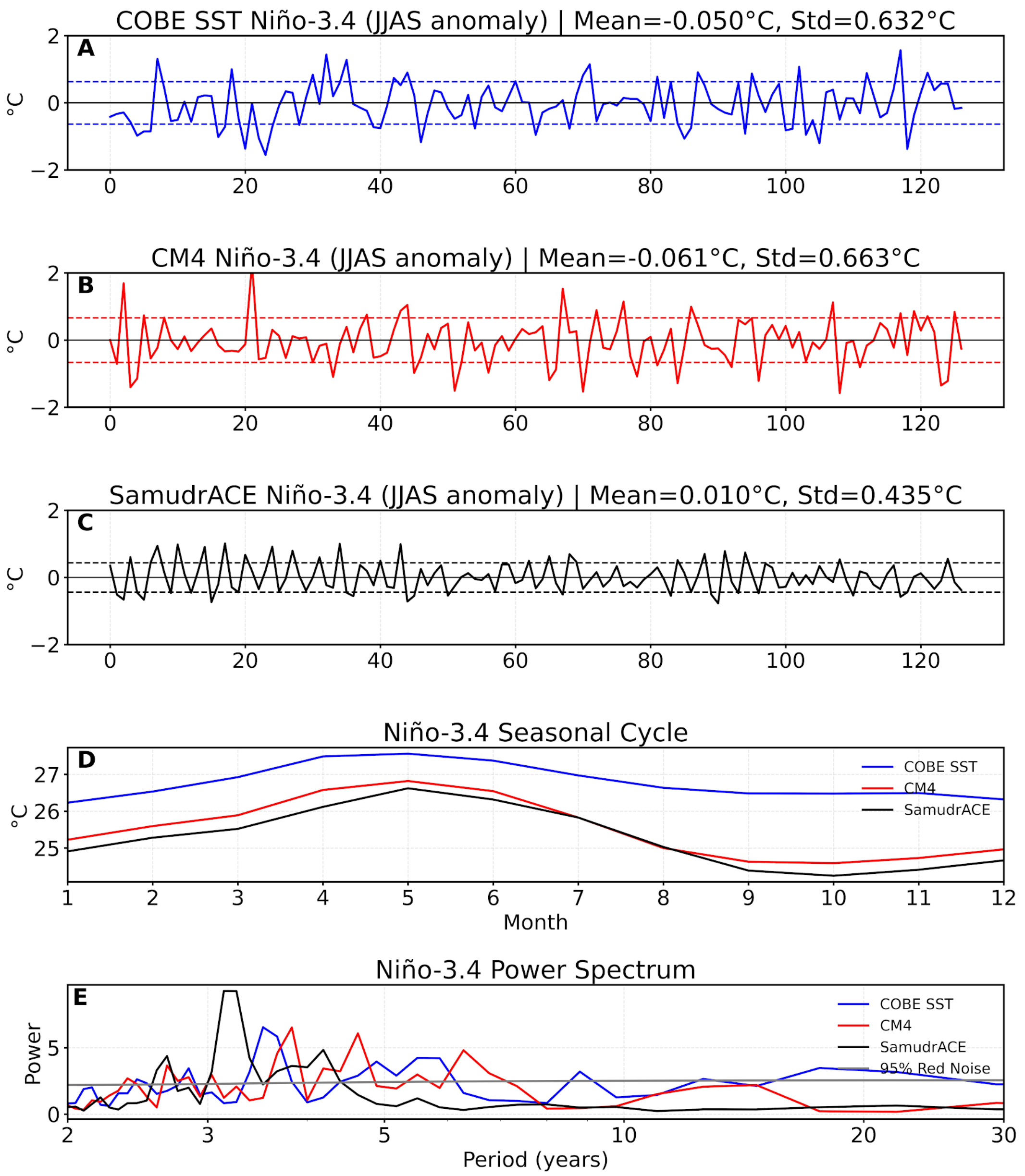

*Figure10. Niño-3.4 sea surface temperature (SST) variability from observations, NOAA Geophysical Fluid Dynamics Laboratory CM4, and SamudrACE over a 126-year period. Panel (A) shows the JJAS mean Niño-3.4 SST anomalies from COBE-SST observations, panel (B) shows the corresponding anomalies from GFDL-CM4, and panel (C) shows the anomalies from SamudrACE. Panel (D) presents the climatological monthly mean Niño-3.4 SST from COBE-SST (blue), GFDL-CM4 (red), and SamudrACE (black). Panel (E) shows the power spectra of monthly Niño-3.4 SST anomalies for the three datasets, highlighting the dominant timescales of ENSO*

by CM4 (Fig.9D). This discrepancy may have implications for simulating the ENSO–monsoon relationship.

To examine the temporal characteristics of ENSO, we construct time series of JJAS mean Niño-3.4 SST anomalies from observations, SamudrACE, and CM4 (Figure 10A–C, respectively). A comparison of time series of Nino3.4 JJAS mean SST anomalies simulated by SamudrACE (Fig.10C) with those from CM4 (Fig.10B) and observations (Fig.10A) show that the ESM emulator fails reproduce the full observed range of ENSO extremes thereby significantly underestimating the amplitude of its interannual variability, with a standard deviation of 0.435°C compared with 0.632°C in observations and 0.663°C in CM4. The other notable bias is that the interannual variability simulated by SamudrACE is too regular compared to either observations or CM4 simulations with a dominant period ~ 3 years (Fig.10E). The annual cycle of climatological mean Niño-3.4 SST (Figure 10D) indicates that SamudrACE exhibits a cold bias of approximately 1.0°C during boreal summer and as large as 1.5°C during boreal winter. This cold bias is also present in CM4, indicating that the mean-state bias in SamudrACE is inherited from its training data. Our estimate of uncertainty in the observed climatological annual cycle of Nino3.4 SST from COBE is ~0.15-0.20 (not shown). Thus the simulated cold bias is highly significant. The power spectra of monthly Niño-3.4 SST anomalies from observations and SamudrACE (Figure 10E) show that ENSO variability in SamudrACE is dominated by a period of about 3 years, whereas observations are dominated by variability around 4 years, with additional power at 6 and 9 years. Thus, although the climatological cold bias in SamudrACE is inherited from CM4, the amplitude of ENSO variability is further reduced in the emulator relative to both CM4 and observations pointing towards some systemic problem of the deep learning model in emulating the unstable air-sea interaction that leads to the amplification of the ENSO. The bias of simulating a too regular ENSO compared to observation has been noted in Duncan et al. (2026) as well (their Fig.4 and Fig.S10).

A simultaneous correlation between ISMR and Niño-3.4 SST is considered a measure of the ENSO-monsoon relationship. A 21-year moving correlation between the two (Figure 11A) shows pronounced epochal variability, with a long-term mean correlation of −0.46 in the observations. SamudrACE reproduces this behavior but with larger epochal fluctuations and a weaker long-term mean correlation of −0.36. Another key feature of the ENSO–monsoon relationship is that the negative correlation peaks not during the peak monsoon months but about three months afterward (Choudhury et al., 2022). The lagged relationship may be considered a potential two-way interaction between ISMR and the ENSO (Kirtman and Shukla, 2000; Webster and Yang 1992). To assess the ability of SamudrACE to simulate this behavior, we examine lead–lag correlations between ISMR and two ENSO indices: monthly Niño-3.4 SST (Figure 11C) and a global-ENSO index (Figure 11B). CM4 reproduces the observed phase relationship reasonably well, with the maximum negative correlation occurring when ENSO lags ISMR by approximately three months. In contrast, SamudrACE fails to capture this timing and simulates the maximum negative correlation when ISMR leads by about 12 months with respect to both Nino3.4 SST or a global-ENSO index. This seems to be consistent with the significantly weaker interannual variability of Nino3.4 SST simulated by SamudrACE and its

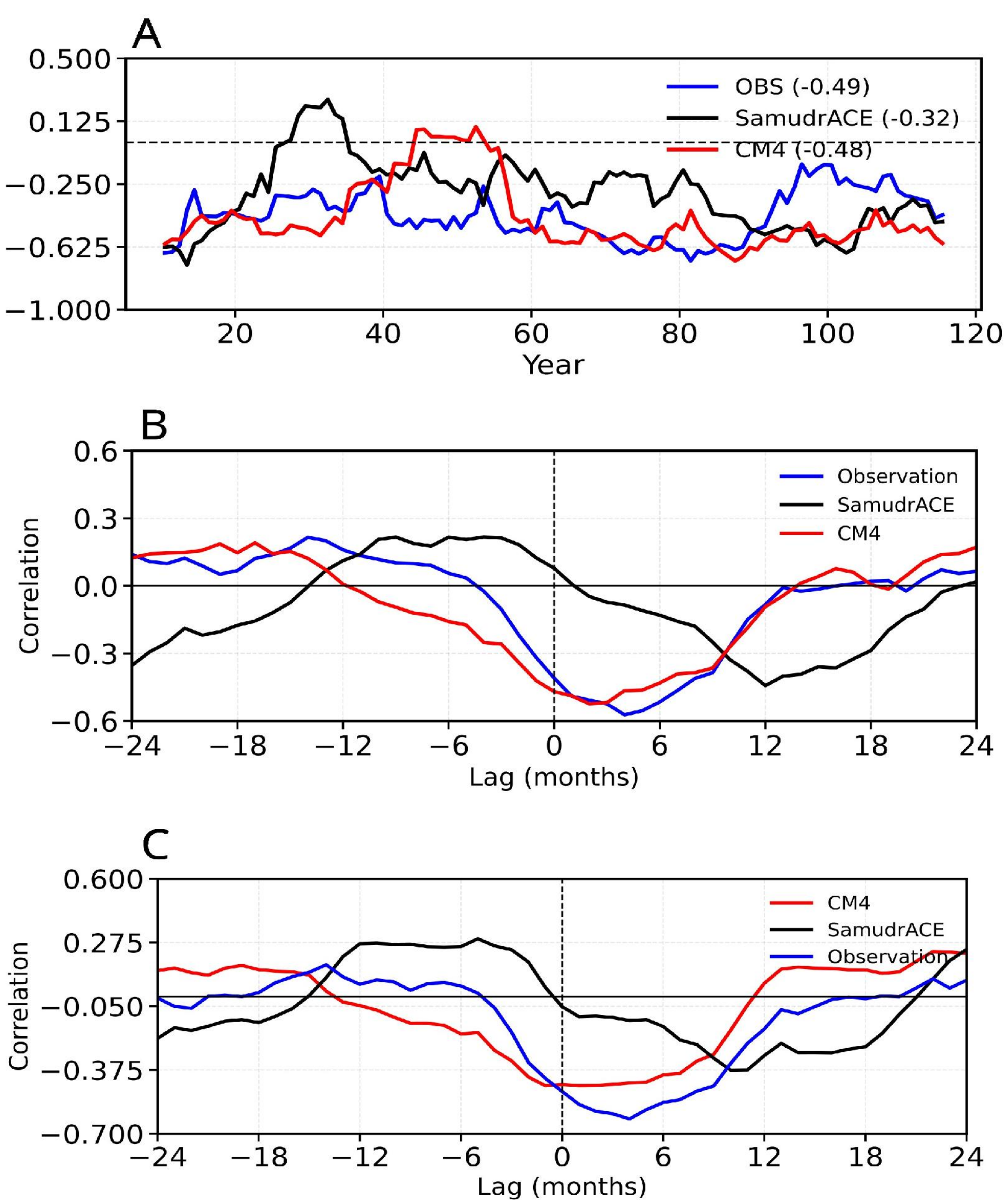

*Figure11 (A) 21-year moving correlation between Niño-3.4 SST anomalies and JJAS Indian monsoon rainfall for observations, with July as the center month, along with the corresponding 21-year moving correlation from SamudrACE simulations. (B) Lead–lag correlation between JJAS Indian monsoon rainfall and a global ENSO index, defined as the first principal component (PC1) of monthly SST anomalies over 30°S–30°N, for both observations and SamudrACE simulations. (C) Lead–lag correlation between simulated monthly Niño-3.4 SST anomalies and simulated JJAS Indian summer monsoon rainfall (ISMR), with July taken as the reference month.*

regular periodic behavior with respect to observations. Because ENSO is the dominant driver of ISMR interannual variability, this incorrect phase relationship may lead to substantial errors in ISMR prediction by SamudrACE.

### 3.3.3 Simulation of large-scale air-sea interaction signature during a Super El Nino

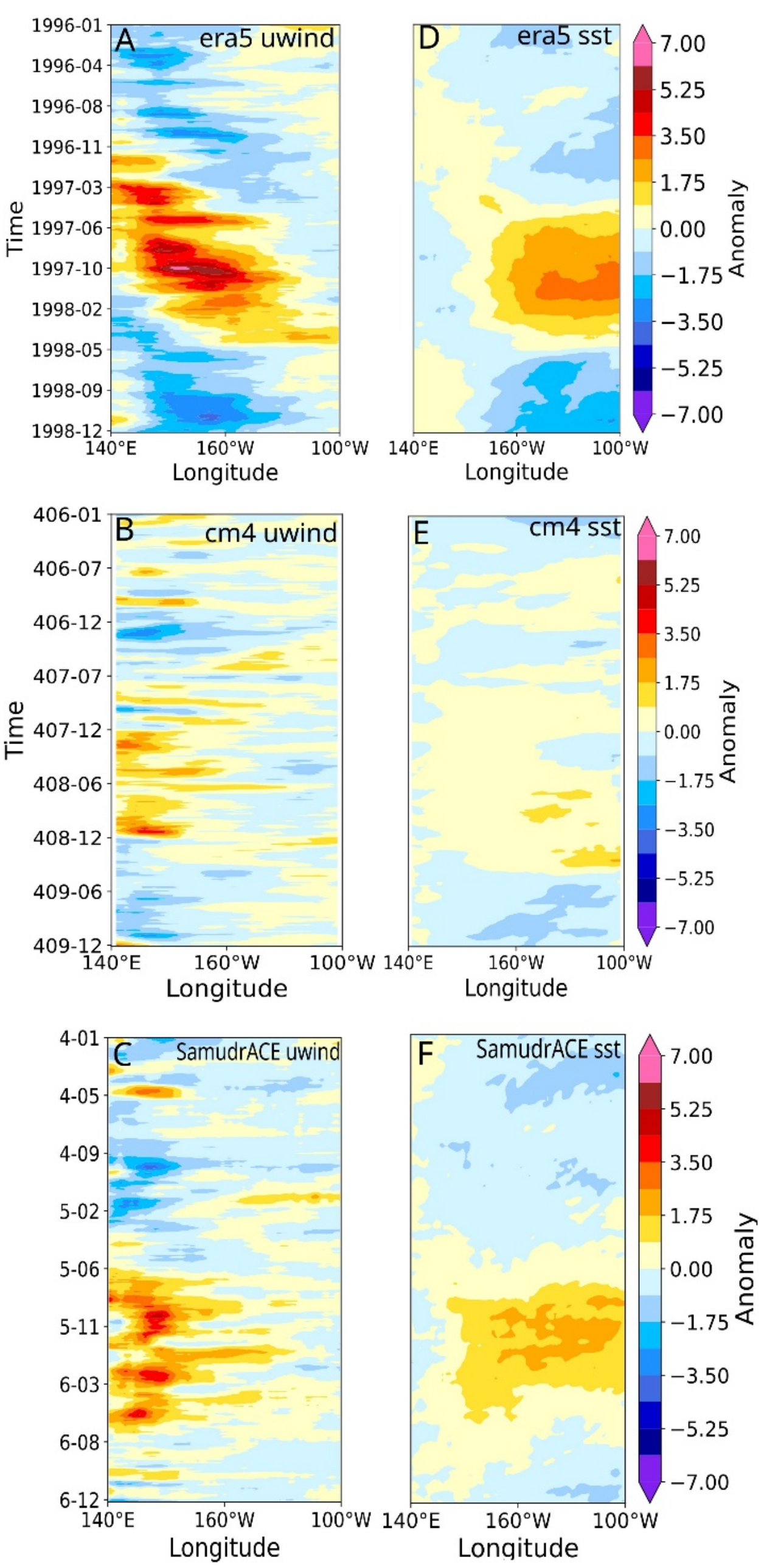

*Figure 12 Time–longitude evolution over the Pacific basin (140°E–100°W) of equatorial variability (5°S–5°N averaged). Panels (A) and (D) show zonal wind anomalies and sea surface temperature (SST), respectively, from ERA5 for the period 1995–01 to 1998–12. Panels (B) and (E) show the corresponding zonal wind anomalies and SST from CM4 for a 3-year period (406–1 to 409–12). Panels (C) and (F) show zonal wind and SST anomalies for a 3-year period (4–1 to 6–12) from SamudrACE )respectively.*

The ocean–atmosphere interaction associated with ENSO lies at the core of interannual variability of ISMR as well as other tropical climates. A useful test for an Earth system model emulator such as SamudrACE is whether it can realistically simulate the equatorial ocean dynamics associated with air–sea interactions. For this purpose, we examine the evolution of zonal wind anomalies, SST anomalies, and total SST averaged over 5°S–5°N simulated by SamudrACE for a three year period around a super El Niño event from the long simulation (Figure 12B, E), and compare it with the observed evolution during the 1997–1998 super El Niño (Figure 12A, D).

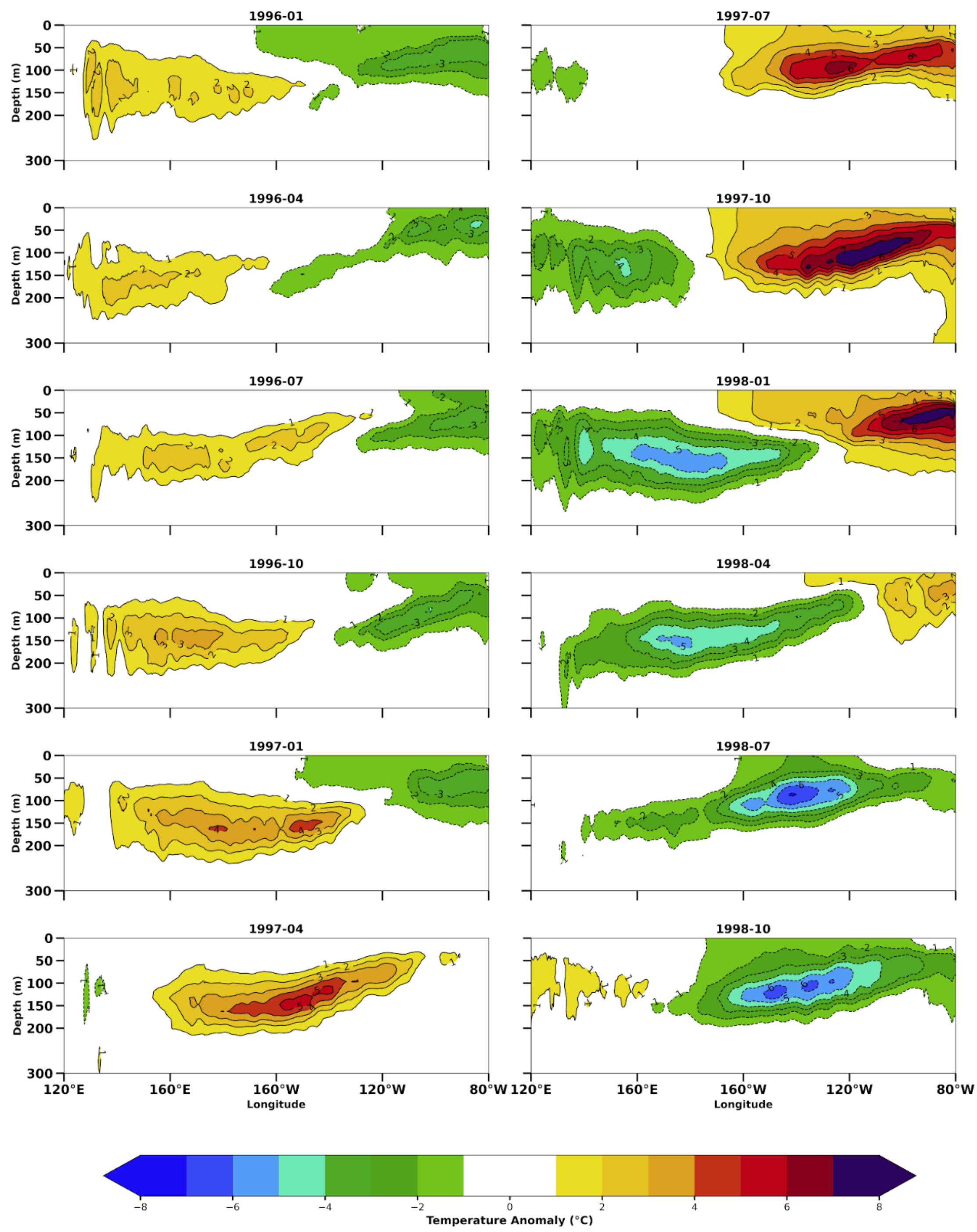

*Figure 13. Longitude–depth sections of equatorial Pacific temperature anomalies (°C) averaged over 5°S–5°N from ORAS5 during 1996–1998.*

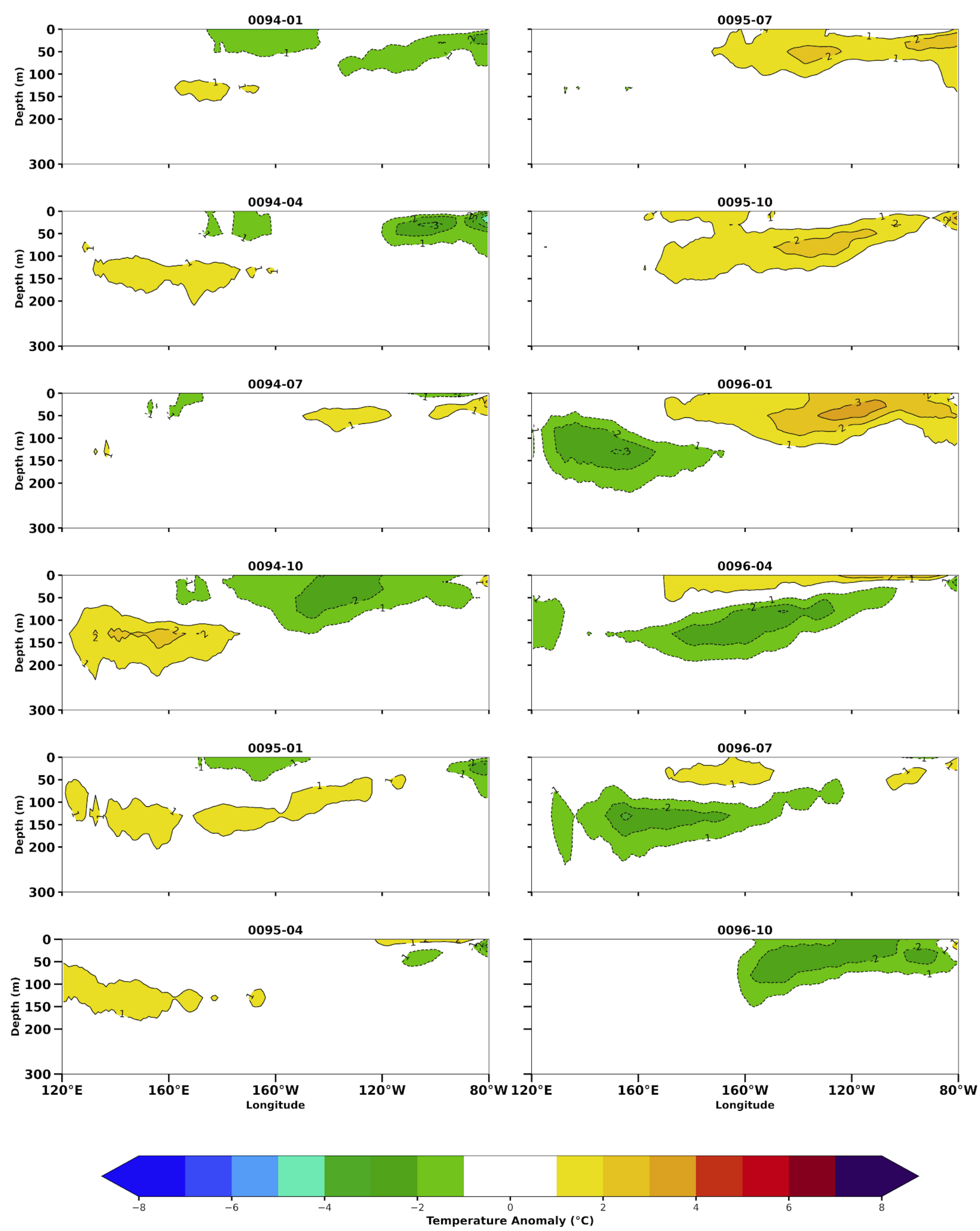

*Figure 14 Longitude–depth sections of equatorial Pacific temperature anomalies (°C) averaged over 5°S–5°N from SamudrACE during 0094-0096 years.*

In both observations (Fig.12A,D) and CM4 (Fig.12E, F) the westerly wind bursts initiate an oceanic Kelvin wave that propagate eastward taking westerly zonal wind anomalies at the wake of it through damped air-sea interactions. Such streaks of westerly wind anomalies can be seen propagating eastward at the oceanic Kelvin wave speed in all simulations including that by SamudrACE (Fig.12B) without amplification, The preconditioning of the equatorial thermocline by the initial Kelvin waves trigger a unstable air-sea interaction (Philander et al., 1984) and starts amplifying both westerly zonal winds to the west and SST anomalies to the east. The unstable air-sea interaction is associated with a 'slow' thermocline wave propagating eastwards making westerly bursts to be more persistent and helping the SST anomalies to amplify. While the process is working in CM4 (Fig.12E, F), the 'super' El Nino in SamudrACE (Fig.12B,E) is highly subdued compared to CM4 indicating a problem with the deep learning model in amplifying the unstable air-sea interaction. Thus, although SamudrACE simulates super El Niño-like events, further analysis is required to determine whether these events arise from physically consistent processes.

To complement the biases in surface manifestation of the air-sea interaction, we also examine the slow evolution of the thermocline wave during such a super-El Nino event. For this purpose, longitude - depth cross section of evolution of temperature anomalies for three-year period around the super-El Nino are shown in Fig.13 and Fig. 14 from observations (ORAS5, Zuo et al., 2019) and SamudrACE respectively. In observations, a seed for the super-El Nino (1998-01) was seen as a sub-surface warm water anomaly in western Pacific nearly two years ago at a depth between 130m-110m between 130oE-170oE. This warm water anomaly amplifies as it propagates eastwards and outcrops in the eastern Pacific establishing the El Nino by 1997-07. At this time a weak cold subsurface water anomaly appeared in the western Pacific as seed for the next La Nina event. The sub-surface cold-water anomaly amplifies as it propagated eastwards along the thermocline, eventually outcropping in 1998-07 and initiating the surface manifestation of a La Nina (Fig.13). As seen in the surface (Fig.12), the timing of initiation of the seed thermocline wave and its eastward propagation in SamudrACE (Fig.14) is comparable to observations (Fig.13) but the rate of amplification is much slower leading to a much weaker El Nino or La Nina compared to observation. However, while there is a coupled instability even in SamudrACE, why is the air-interaction unable to amplify requires further analysis and sensitivity experiments. A notable mean bias exists in the climatological D20 structure (Fig. S5). In the western Pacific (around 120oE), the observed D20 is approximately 160 m, whereas SamudrACE simulates a shallower thermocline of about 110 m. In the eastern Pacific (around 70°W), where the observed D20 is approximately 40 m, the model places the 20°C isotherm close to the surface. These biases may influence the simulation of tropical air-sea interaction and the coupled instability.

**Discussions**

To explore the potential of an Earth system model emulator for accurate and cost effective S2S prediction of Indian monsoon rainfall, we examine the fidelity of SamudrACE, the first successful coupled emulator, in simulating the observed intraseasonal and interannual variability of Indian monsoon rainfall. We do not claim this analysis to be either rigorous or exhaustive. To fast track the development of a cost effective and skillful deep learning S2S prediction system for Indian monsoon, our objective is to obtain a first-order estimate of the major biases of the emulator and to identify directions for improving the simulation of intraseasonal and interannual variability of ISMR. Considering the pre-eminent state of SamudrACE in the evolution of deep learning ESM development, biases are expected. While our work (analysis of the biases) may not be fundamental in nature, it is relevant and important for accelerating development of powerful deep learning S2S prediction systems for societal benefit.

To test the robustness of the variability simulated by SamudrACE, we generated three ensemble members of 300-year pre-industrial simulations. To examine whether any climate drift exists in the simulation of the Indian monsoon, we analyze time series of JJAS rainfall over Indian land points from all three ensemble members (Figure S1). While multidecadal natural variability in ISMR is expected in the simulations, a long-term trend is not. Among the three ensemble simulations, two (Figure S1A, C) do not show statistically significant trends, while one (Figure S1B) shows a weak but statistically significant decreasing trend. The amplitude of interannual variability in each ensemble is comparable, approximately 62.0 mm ± 2.5 mm. All ensemble members exhibit significant natural multidecadal variability (Figure S2E) but no long-term trends as expected. Each ensemble shows a prominent multidecadal mode with a period of about 60 to 80 years, consistent with observed ISMR variability (Rajesh & Goswami, 2020). Such variability in SamudrACE likely arises from internal ocean atmosphere feedback. Therefore, from the perspective of developing an S2S prediction system, the ability of SamudrACE to simulate realistic internal multidecadal variability is a strength.

In terms of weaknesses, (i) the first major weakness that both CM4 and SamudrACE has is a large cold bias (Fig.1) in simulating not only Nino3.4 SST but global SST (Fig.S4) except over some coastal regions where they have a warm bias. (ii) The second important bias is in simulating the ENSO phenomenon. The average amplitude of ENSO simulated by SamudrACE is about half of that observed and the weak ENSO like events are too periodic with a dominant ~3-year periodicity devoid of the intermittency inherent in observed ENSO or simulated by CM4. (iii) The third bias important for seasonal prediction of Indian monsoon is the poor simulation of lead-lag relationship between ISMR and Nino3.4 SST. In contrast to observation and in CM4 the correlation between ISMR and Nino3.4 SST is strongest after the monsoon season with ISMR leading by 3-months, the SamudrACE simulates it 12-months after peak monsoon. It is possible that all three major biases are interlinked through biases in simulating air-sea interactions due to weak oceanic Kelvin waves (as seen in the atmosphere, Fig. 1) or through the coupling strategy. However, it is noted that the first two major biases come directly from training while the third bias is intrinsic to the deep learning model.

Most biases in SamudrACE are linked to its training data from CM4. Therefore, there is substantial scope for improving S2S simulations by retraining the emulator using coupled ocean atmosphere model outputs that better represent Indian monsoon variability. We have identified about a dozen relatively high resolution CMIP6 models (Choudhury et al., 2022) with lower biases in simulating the mean and interannual variability of Indian monsoon rainfall. We propose retraining SamudrACE using simulations from ten CMIP6 models, each with ten ensemble members. A potential problem may be availability of sub-daily data from pre-industrial simulations of the CMIP6 models.

We also recognize that the weak amplitude of ENSO in SamudrACE compared to observations is a result of its inability to amplify the coupled instability associated with the ENSO. The subsurface ocean dynamics is likely to play an important role in the teleconnection between ENSO and the Indian monsoon (Sharma et al., 2022, 2025) as well as in amplifying the ocean-atmosphere interaction. In this context, adequate vertical resolution in the upper 300 m of the ocean may be critical for accurately simulating monsoon rainfall variability. We therefore propose redesigning the ocean component of SamudrACE with an appropriate number of vertical levels and improved vertical distribution. However, as indicated in Figure 12, 13 and 14 some biases may arise from the coupling strategy between the ocean and atmospheric components. In this context, it may be noted that while the atmosphere is trained on 6-hourly analysis, the SamudrACE ocean component is trained and coupled using 5-day mean fields. It is possible that weak ENSO simulation in SamudrACE is associated with the coupling strategy. There is some indication that higher frequency of coupling every 12 hours may improve ENSO amplitude simulation in a coupled emulator (Wang et al., 2024). We propose investigating this issue through additional analyses and sensitivity experiments, followed by a redesigning of the coupling strategy based on these findings.

The SamudrACE is in its 'infant' stage of development as far as deep learning ESM development goes. One strategy to significantly reduce the major biases may be to train the model on high resolution CMIP6 model simulations and fine tune the training through transfer learning using observations. This has worked in long-lead seasonal prediction of ISMR (Sharma et al., 2026). We plan to use this strategy on all the development plans indicated above. Our analysis raises some interesting questions for further in-depth examination of the SamudrACE simulations themselves. While biases in tropical air-sea interactions practically kills the dominant tropical variability namely the ENSO, SamdrACE surprisingly simulates realistic large amplitude multi-decadal natural oscillation of ISMR making it a good candidate for decadal prediction of ISMR. In observations, the ISMR multi-decadal variability is related to Atlantic multi-decadal variability which in turn is linked to the AMOC (Rajesh and Goswami, 2020). It is notable that the power spectrum of 300-year simulation of JJAS mean ISMR by CM4 (Fig.S2) does not have a statistically significant multi-decadal variability with period ~70 years. Thus, SamudrACE is not learning it from training. How does the coupled emulator achieve this rather difficult task even when it fails to emulate the tropical air-sea interactions with fidelity remains an open question requiring a good deal of additional investigation outside the scope of the present study.

**Acknowledgements:** BNG thanks Anusandhan National Research Foundation, Government of India for support through the ANRF Prime Minister Fellowship and Gauhati University for hosting. BKB acknowledges the Department of Instrumentation and USIC and ST radar Centre for the Project Associate Program. BKB and US acknowledge Gauhati University for providing support to carry out the research work. DS and RIS acknowledge the funding from IndusInd Bank through the CSR Project (CR23242519AEINIB002696) and Institute of Eminence (IoE) initiative of the Ministry of Human Resources and Development (MHRD) at IIT Madras (Grant No. SP22231222CPETWOCTSHOC). SKS thanks MoES, Government of India for all the support.

**Supplementary Files**

**For**

# Deep Learning Earth System Model Simulation of Indian Monsoon Intraseasonal and Interannual Variability

Bijit Kumar Banerjee[1,2], Devabrat Sharma[3,4], R. I. Sujith[3,4], Chandrashekar Lakshminarayanan[5], Manikandan Narayanan[5], Subodh K. Saha[6], Anurag Dipankar[7], Utpal Sarma[1,2], B. N. Goswami[1]

[1]ST Radar Centre, Gauhati University, India

[2]Department of Instrumentation and USIC

[3]Department of Aerospace Engineering, Indian Institute of Technology Madras, Chennai-600036, India

[4]Centre of Excellence for Studying Critical Transitions in Complex Systems, Indian Institute of Technology Madras, Chennai-600036, India.

[5]Department of Computer Science and Engineering, Indian Institute of Technology Madras, Chennai, India

[6]Indian Institute of Tropical Meteorology, Pune, India

[7]Institute for Atmospheric and Climate Science, ETH Zürich

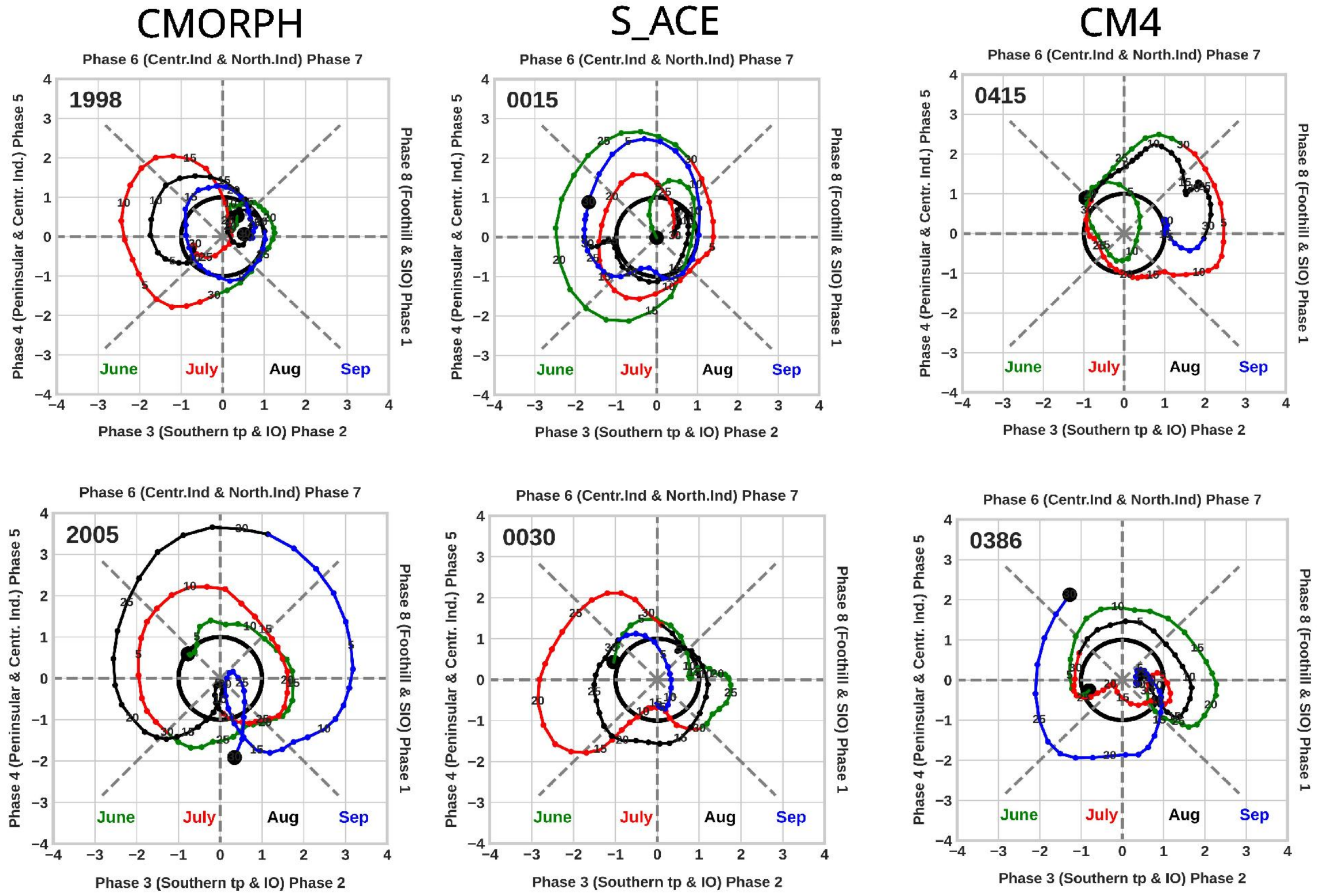

*Figure S1: Illustration of real time monitoring or evolution of the MISO. The location in the phase space of MISO1 and MISO2 on a given day indicates one of the eight phases of the MISO. Each quasi-circular episode is an individual event, and the size of the quasi-circle indicates it amplitude. Here, just two years are shown for illustration. However, evolution cannot be compared directly as SamudrACE simulations do not map directly to either CM4 and observations.*

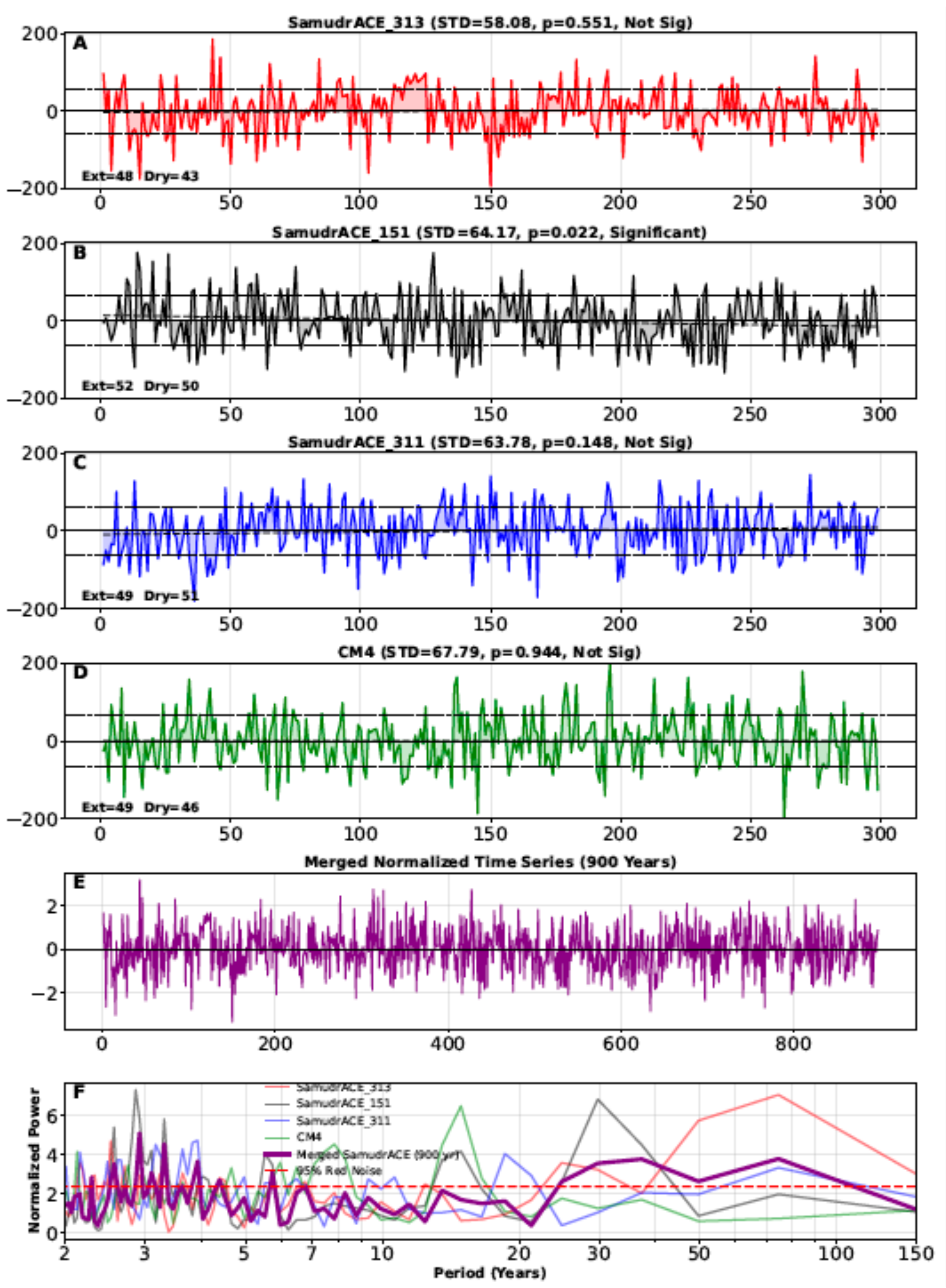

*Figure S2. (A–C) Time series of rainfall over the Indian landmass from three different ensemble members of SamudrACE, each initialized in different years and forced identically. **(D)** Time series of rainfall over the Indian landmass from the 300-year CM4 simulation. **(E)** The three SamudrACE simulations are normalized by their respective standard deviations and concatenated to construct a continuous 900-year time series. **(F)** Power spectra of the three individual SamudrACE simulations, the 300-year CM4 simulation, and the merged normalized 900-year SamudrACE time series, showing variability at interannual timescales of approximately 3–7 years and multidecadal variability at approximately 30 and 100 years. Overall, all simulations reasonably capture the high-frequency oscillations.*

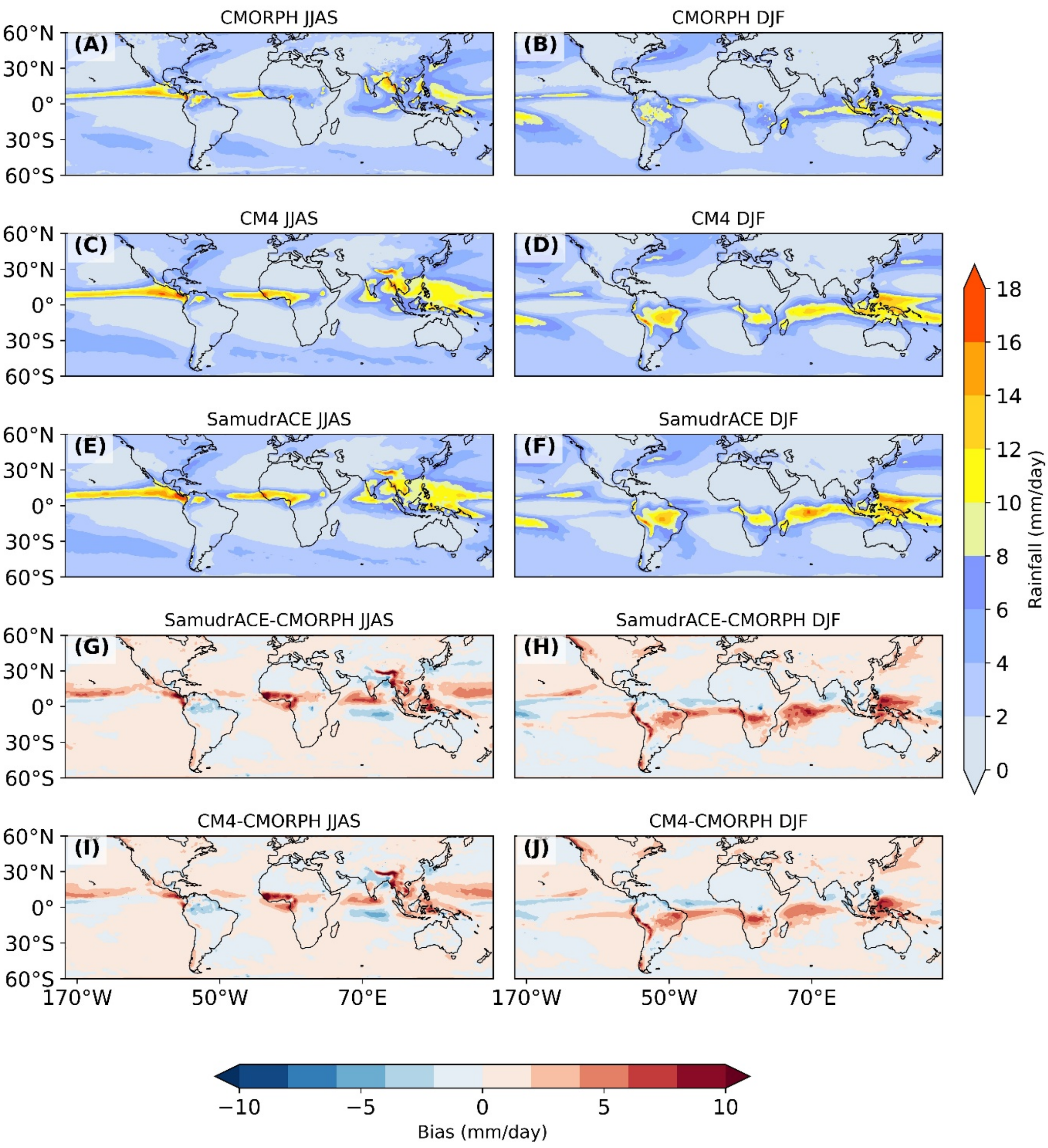

*Figure S4 Global Climatological spatial structure of rainfall for the JJAS (left column) and DJF (right column) seasons. Panels (A, B) show CMORPH; (C, D) show observations from CM4 (100 km); and (E, F) show SamudrACE. Panels (G, H) present the difference between SamudrACE and CMORPH for JJAS and DJF, respectively, while panels (I, J) show the difference between CM4 and CMORPH for JJAS and DJF.*

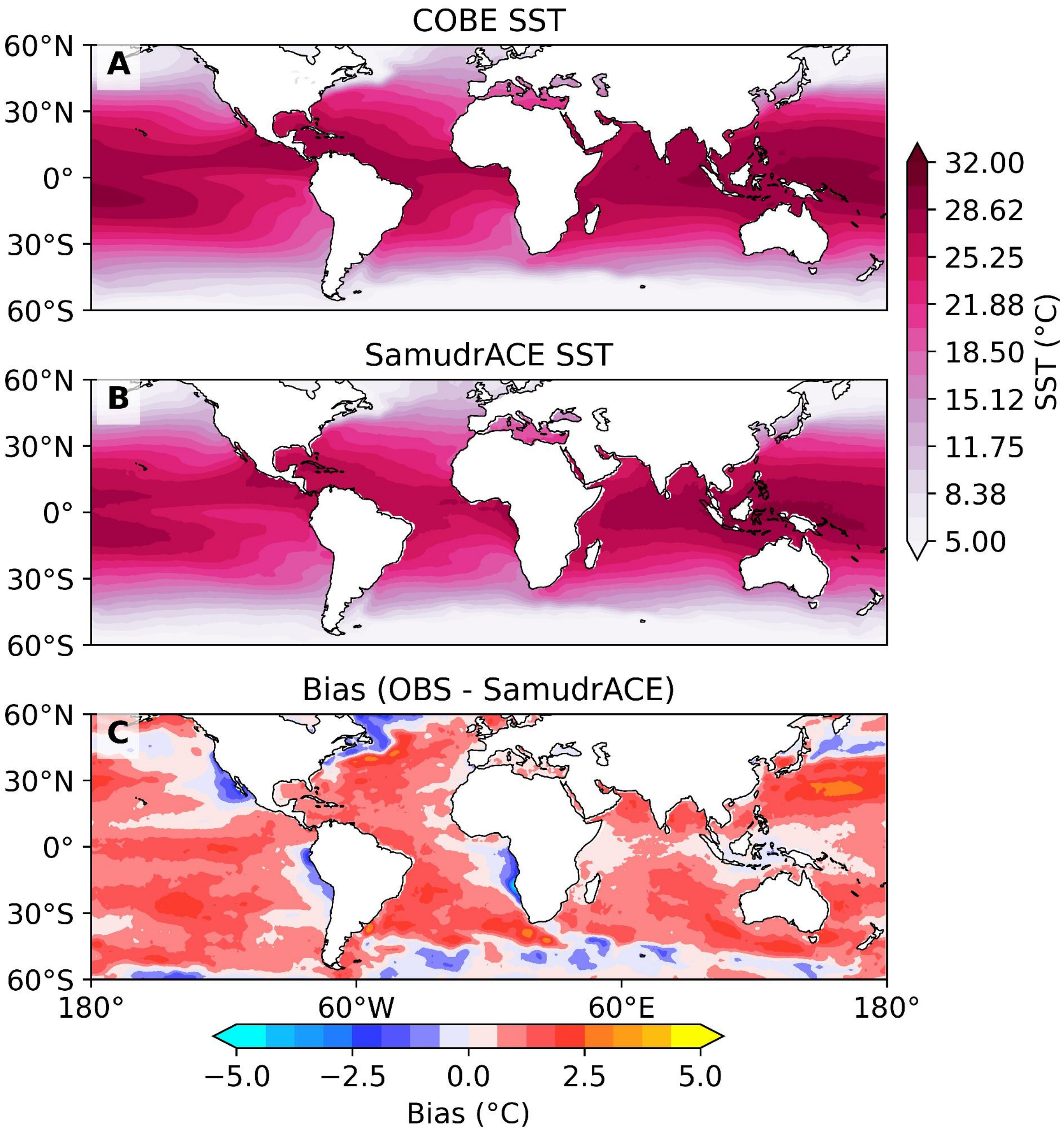

*Figure S4 (A) Annual climatological mean sea surface temperature (SST) over the globe from COBE SST. (B) Same as (A), but for SamudrACE. (C) Bias in SST (observations minus model). SamudrACE exhibits a cold bias in SST.*

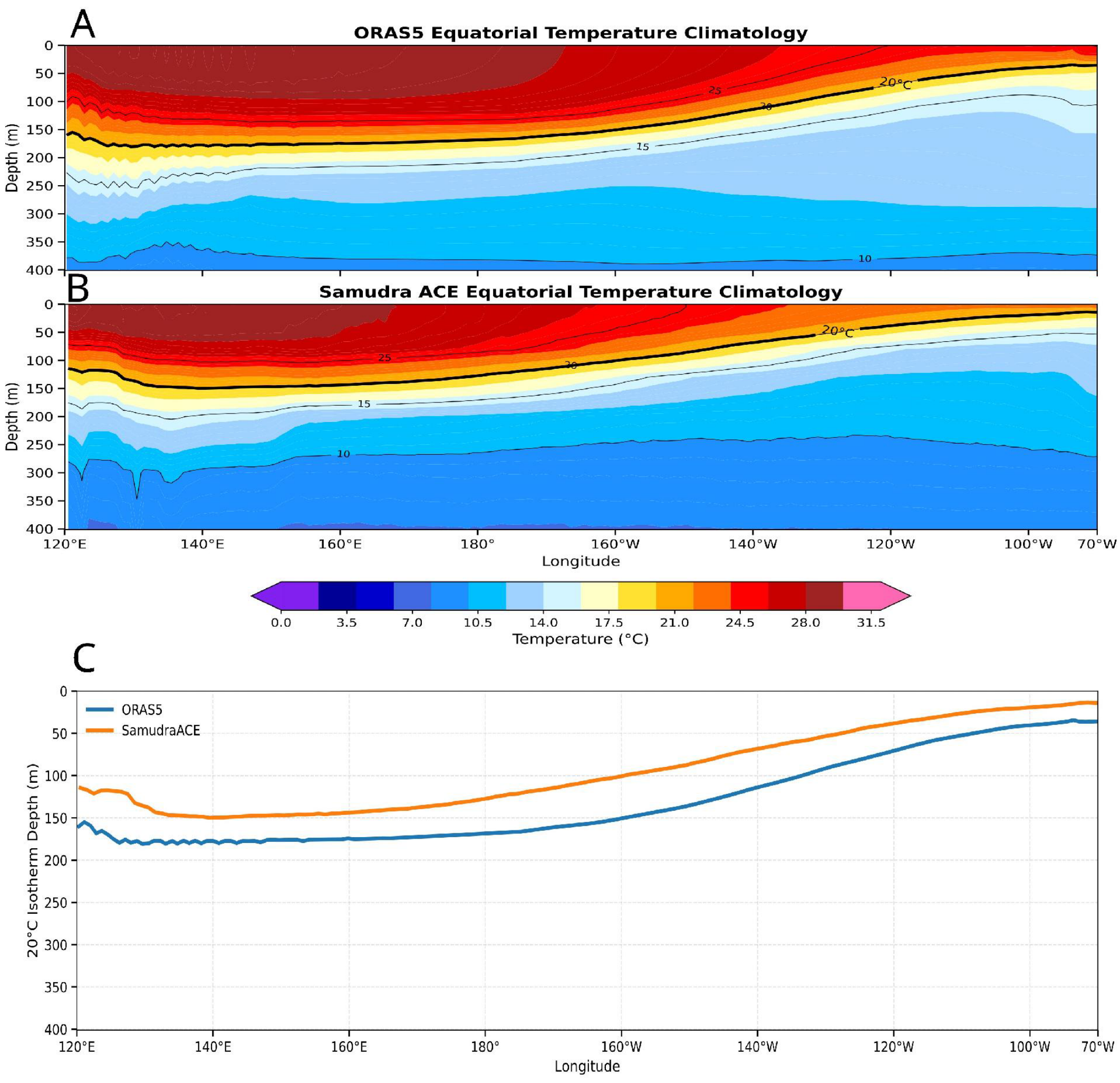

*Figure S5 Twenty-five-years mean equatorial Pacific temperature climatology (°C) from (A) ORAS5 and (B) SamudrACE. Shading shows temperature, black contours are plotted every 5°C, and the thick black contour denotes the 20°C isotherm. (C) Comparison of the 20°C isotherm depth along the equatorial Pacific.*